\documentclass[11pt]{article} 
\usepackage{amsfonts,amssymb,amsmath,amsthm,booktabs,dashrule,dsfont,float,graphicx,hyperref,mathrsfs,euscript,enumitem,bm,bbm,pifont,multirow}
\usepackage[onehalfspacing]{setspace}
\usepackage{caption,subcaption,lscape}
\hypersetup{pdfborder = {0 0 0},colorlinks=true,linkcolor=blue,urlcolor=blue,citecolor=blue}
\usepackage{natbib}
\usepackage[title]{appendix}
\usepackage[top=1in,bottom=1in,left=1in,right=1in]{geometry}
\usepackage{placeins}
\usepackage{threeparttable}
\usepackage[dvipsnames]{xcolor}
\usepackage{cancel,enumitem,array,lscape}
\usepackage[font=small,labelfont=bf,textfont=it]{caption}   
\usepackage{textcomp} 

\usepackage[normalem]{ulem}

\numberwithin{equation}{section}

\definecolor{carmine}{rgb}{0.59, 0.0, 0.09}
\hypersetup{colorlinks=true,linkcolor=carmine,filecolor=magenta,urlcolor=cyan,citecolor=carmine}

\newcounter{example}[section]

\usepackage{listings}
\definecolor{codegreen}{rgb}{0,0.6,0}
\definecolor{codegray}{rgb}{0.5,0.5,0.5}
\definecolor{codepurple}{rgb}{0.58,0,0.82}
\definecolor{backcolour}{rgb}{0.95,0.95,0.95}
 
\lstdefinestyle{Rstyle}{
	language=R,
    backgroundcolor=\color{backcolour},   
    commentstyle=\color{codegreen},
    numberstyle=\tiny\color{codegray},
    stringstyle=\color{codepurple},
    basicstyle=\footnotesize\ttfamily,
    breakatwhitespace=false,         
    breaklines=true,                 
    captionpos=b,                    
    keepspaces=true,                 
    numbers=none,                    
    numbersep=5pt,                  
    showspaces=false,                
    showstringspaces=false,
    showtabs=false,                  
    tabsize=2,
    upquote=true
}
\input{stata-lstlisting}                                

\definecolor{codegreen2}{rgb}{0,0.6,0}
\definecolor{codegray2}{rgb}{0.5,0.5,0.5}
\definecolor{codepurple2}{rgb}{0.58,0,0.82}
\definecolor{backcolour2}{rgb}{0.95,0.95,0.92}
 
\lstdefinestyle{Pystyle}{
    language = Python,
    backgroundcolor=\color{backcolour2},   
    commentstyle=\color{codegreen2},
    keywordstyle=\color{black},
    numberstyle=\tiny\color{codegray2},
    stringstyle=\color{codepurple2},
    basicstyle=\footnotesize,
    breakatwhitespace=false,         
    breaklines=true,                 
    captionpos=b,
    deletekeywords=[2]{id},
    keepspaces=true,                 
    numbers=left,                    
    numbersep=5pt,                  
    showspaces=false,                
    showstringspaces=false,
    showtabs=false,                  
    tabsize=2,
    upquote=true
}

\theoremstyle{definition}
\newtheorem{remark_tmp}{Remark}

\theoremstyle{definition}

\allowdisplaybreaks[1]

\renewcommand*{\arraystretch}{.6}

	\DeclareMathOperator*{\argmin}{arg\,min}
	
	\DeclareMathOperator*{\diag}{diag}

\renewcommand{\P}{\mathbb{P}}
\newcommand{\E}{\mathbb{E}}
\newcommand{\V}{\mathbb{V}}

\newcommand{\I}{\mathds{1}}

\newcommand{\bA}{\mathbf{A}}
\newcommand{\bB}{\mathbf{B}}
\newcommand{\bC}{\mathbf{C}}
\newcommand{\bD}{\mathbf{D}}

\newcommand{\bG}{\mathbf{G}}

\newcommand{\bI}{\mathbf{I}}

\newcommand{\bL}{\mathbf{L}}

\newcommand{\bP}{\mathbf{P}}
\newcommand{\bQ}{\mathbf{Q}}

\newcommand{\bV}{\mathbf{V}}

\newcommand{\bZ}{\mathbf{Z}}

\newcommand{\ba}{\mathbf{a}}

\newcommand{\bg}{\mathbf{g}}
\newcommand{\be}{\mathbf{e}}

\newcommand{\bp}{\mathbf{p}}

\newcommand{\br}{\mathbf{r}}

\newcommand{\bu}{\mathbf{u}}

\newcommand{\bx}{\mathbf{x}}

\newcommand{\bw}{\mathbf{w}}

\newcommand{\ttb}{\star}

\newcommand{\bbeta}{\boldsymbol{\beta}}

\newcommand{\bgamma}{\boldsymbol{\gamma}}

\newcommand{\bSigma}{\boldsymbol{\Sigma}}

\newcommand{\blambda}{\boldsymbol{\lambda}}
\newcommand{\bdelta}{\boldsymbol{\delta}}

\newcommand{\teq}{\mathtt{eq}}
\newcommand{\tin}{\mathtt{in}}

\usepackage{listings}
\begin{document}

\title{\texttt{scpi}: Uncertainty Quantification for Synthetic Control Methods\bigskip}
\author{Matias D. Cattaneo\thanks{Department of Operations Research and Financial Engineering, Princeton University.} \and
	    Yingjie Feng\thanks{School of Economics and Management, Tsinghua University.}  \and
	    Filippo Palomba\thanks{Department of Economics, Princeton University.}  \and
	    Rocio Titiunik\thanks{Department of Politics, Princeton University.}}
\maketitle

\begin{abstract}
	The synthetic control method offers a way to quantify the effect of an intervention using weighted averages of untreated units to approximate the counterfactual outcome that the treated unit(s) would have experienced in the absence of the intervention. This method is useful for program evaluation and causal inference in observational studies. We introduce the software package \texttt{scpi} for prediction and inference using synthetic controls, implemented in \texttt{Python}, \texttt{R}, and \texttt{Stata}. For point estimation or prediction of treatment effects, the package offers an array of (possibly penalized) approaches leveraging the latest optimization methods. For uncertainty quantification, the package offers the prediction interval methods introduced by \citet*{Cattaneo-Feng-Titiunik_2021_JASA} and \citet*{Cattaneo-Feng-Palomba-Titiunik_2022_wp}. The paper includes numerical illustrations and a comparison with other synthetic control software.
	
\end{abstract}


\thispagestyle{empty}
\clearpage

\doublespacing
\setcounter{page}{1}

\pagebreak
\setcounter{page}{0}\thispagestyle{empty}
\tableofcontents
\pagebreak

\pagestyle{plain}

\section{Introduction}

The synthetic control method was introduced by \citet*{Abadie-Gardeazabal_2003_AER}, and since then it has become a popular approach for program evaluation and causal inference in observational studies. It offers a way to study the effect of an intervention (e.g., treatments at the level of aggregate units, such as cities, states, or countries) by constructing weighted averages of untreated units to approximate the counterfactual outcome that the treated unit(s) would have experienced in the absence of the intervention. While originally developed for the special case of a single treated unit and a few control units over a short time span, the methodology has been extended in recent years to a variety of other settings with longitudinal data. See \citet*{Abadie_2021_JEL} for a review on synthetic control methods, and \citet*{Abadie-Cattaneo_2018_ARE} for a review on general methods for program evaluation.

Most methodological developments in the synthetic control literature have focused on either expanding the causal framework or developing new implementations for prediction/point estimation. Examples of the former include disaggregated data settings \citep*{Abadie-LHour_2021_JASA} and staggered treatment adoption \citep*{BenMichael-Feller-Rothstein_2022_JRSSB}, while examples of the latter include employing different constrained estimation methods (see Table \ref{tab:lit_constr} below for references). Conceptually, implementation of the synthetic control method involves two main steps: first, treated units are ``matched'' to control units using only their pre-intervention data via (often constrained) regression methods and, second, prediction of the counterfactual outcomes of the treated units are obtained by combining the pre-intervention ``matching'' weights with the post-intervention data of the control units. As a result, the synthetic control approach offers a prediction or point estimator of the (causal) treatment effect for the treated unit(s) after the intervention was deployed.

Compared to prediction or estimation, considerably less effort has been devoted to develop principled uncertainty quantification for synthetic control methods. The most popular approach in practice is to employ design-based permutation methods taking the potential outcome variables as non-random \citep*{Abadie-Diamond-Hainmueller_2010_JASA}. Other approaches include methods based on large-sample approximations for disaggregated data under correctly specified factor-type models \citep*{Li_2020_JASA}, time-series permutation-based inference \citep*{Chernozhukov-Wuthrich-Zhu_2021_JASA}, large-sample approximations for high-dimensional penalization methods \citep*{Masini-Medeiros_2021_JASA}, and cross-sectional permutation-based inference in semiparametric duration-type settings \citep{Shaikh-Toulis_2021_JASA}. A conceptually distinct approach to uncertainty quantification is proposed by \citet*{Cattaneo-Feng-Titiunik_2021_JASA} and \citet*{Cattaneo-Feng-Palomba-Titiunik_2022_wp}, who take the potential outcome variables as random and develop prediction intervals for the imputed (counterfactual) outcome of the treated unit(s) in the post-intervention period employing finite-sample probability concentration methods.

This article introduces the software package \texttt{scpi} for prediction and inference using synthetic control methods, implemented in \texttt{Python}, \texttt{R}, and \texttt{Stata}. For prediction or point estimation of treatment effects, the package offers an array of possibly penalized approaches leveraging the latest conic optimization methods \citep*{Domahidi-Chu-Boyd_2013_ECOS, Fu-Narasimhan-Boyd_2020_JSS}; see also \citet{Boyd_2004_BOOK} for an introduction. For uncertainty quantification, the package focuses on the aforementioned prediction interval methods under random potential outcomes. The rest of the article focuses on the $\mathtt{R}$ implementation of the software, but we briefly illustrate analogous functionalities for \texttt{Python} in Appendix \ref{app:python}, and for \texttt{Stata} in Appendix \ref{app:stata}.

The \texttt{R} package \texttt{scpi} includes the following six functions:
\begin{itemize}[leftmargin=*]
    \item \texttt{scdata()} and \texttt{scdataMulti()}. These functions take as input a \texttt{DataFrame} object and process it to prepare the data matrices used for point estimation/prediction and inference/uncertainty quantification.  The function \texttt{scdata()} is specific to the single treated unit case, whereas \texttt{scdataMulti()} can be used with multiple treated units and/or when treatment is adopted in a staggered fashion. Both functions allow the user to specify multiple features of the treated unit(s) to be matched by the synthetic unit(s), as well as feature-specific covariate adjustment, and can handle both independent and identically distributed (i.i.d.) and non-stationary (cointegrated) data.
    
    \item \texttt{scest()}. This function handles ``\texttt{scpi\_data}'' objects produced with \texttt{scdata()} or ``\texttt{scpi\_data\_multi}'' objects produced with \texttt{scdataMulti()}, and then implements a class of synthetic control predictions/point estimators for quantification of treatment effects. The implementation allows for multiple features, with and without additional covariate adjustment, and for both stationary and non-stationary data. The allowed prediction procedures include unconstrained weighted least squares as well as constrained weighted least squares with simplex, lasso-type, ridge-type parameter space restrictions and combinations thereof (see Table \ref{tab:pop_constr} below).
    
    \item \texttt{scpi()}. This function takes as input an ``\texttt{scpi\_data}'' object produced with \texttt{scdata()} or an ``\texttt{scpi\_data\_multi}'' object produced with \texttt{scdataMulti()}, and then computes prediction intervals for a class of synthetic control predictions/point estimators for quantification of treatment effects. It relies on \texttt{scest()} for point estimation/prediction of treatment effects, and thus inherits the same functionalities of that function. In particular, \texttt{scpi()} is designed to be the main function in applications, offering both predictions/point estimators for treatment effects as well as inference/uncertainty quantification (i.e., prediction intervals) for synthetic control methods. The function also allows the user to model separately in-sample and out-of-sample uncertainty, offering a broad range of options for practice.
    
    \item \texttt{scplot()} and \texttt{scplotMulti()}. These functions process objects whose class is either ``\texttt{scest}'' or ``\texttt{scpi}''. These objects contain the results of the point estimation/prediction or uncertainty quantification methods, respectively. The commands build on the \texttt{ggplot2} package in \texttt{R} \citep{Wickham_2016_book} to compare the time series for the outcome of the treated unit(s) with the outcome time series of the synthetic control unit, along with the associated uncertainty. The functions return a \texttt{ggplot} object that can be further modified by the user.
\end{itemize}
The objects returned by \texttt{scest()} and \texttt{scpi()} support the methods \texttt{print()} and \texttt{summary()}. In typical applications, the user will first prepare the data using the function \texttt{scdata()} or \texttt{scdataMulti()}, and then produce predictions/point estimators for treatment effects with uncertainty quantification using the function \texttt{scpi()}. The function \texttt{scest()} is useful in cases where only predictions/point estimators are of interest. Numerical illustrations are given in Section \ref{sec: illustration}.

There are many \texttt{R}, \texttt{Python}, and \texttt{Stata} packages available for prediction/point estimation and inference using synthetic control methods; Table \ref{tab:comparison} compares them to the package \texttt{scpi}. As shown in the table, \texttt{scpi} is the first package to offer uncertainty quantification using prediction intervals with random potential outcomes for a wide range of different synthetic control predictors. The package is also one of the first to handle multiple treated units and staggered treatment adoption, offering a wider array of options in terms of predictors and inference methods when compared with the other packages currently available. Furthermore, the package includes misspecification-robust methods, employs the latest optimization packages for conic programs available, and offers automatic parallelization in execution whenever multi-core processors are present, leading to significant improvements in numerical stability and computational speed. Finally, \texttt{scpi} is the only package available in \texttt{Python}, \texttt{R}, and \texttt{Stata}, which gives full portability across multiple statistical software and programming languages, and also the only package employing directly native conic optimization via the \texttt{ECOS} solver (see Table \ref{tab:speed comparison} for details). 

\begin{table}[h]
  \centering
  \caption{Comparison of different packages available on \texttt{PyPi}, \texttt{CRAN}, \texttt{REPEC}, or \texttt{GitHub}.}
  \resizebox{1\columnwidth}{!}{
    \begin{tabular}{lcccccccc}
    \toprule
    \multicolumn{1}{l}{Package}     &\multicolumn{1}{l}{Statistical}     &\multicolumn{1}{c}{Prediction} & \multicolumn{1}{c}{Inference} & \multicolumn{1}{c}{Multiple} & \multicolumn{1}{c}{Staggered} & \multicolumn{1}{c}{Misspecification}   & \multicolumn{1}{c}{Automatic}      & \multicolumn{1}{c}{Last} \\
    \multicolumn{1}{l}{Name}  &\multicolumn{1}{c}{Platform}  &\multicolumn{1}{c}{Method}   & \multicolumn{1}{c}{Method} & \multicolumn{1}{c}{Treated}  & \multicolumn{1}{c}{Adoption}  & \multicolumn{1}{c}{Robust}& \multicolumn{1}{c}{Parallelization}    & \multicolumn{1}{c}{Update} \\
        \midrule
    \href{https://cran.r-project.org/web/packages/ArCo/index.html}{\texttt{ArCo}} & \texttt{R} &\texttt{LA} & \texttt{Asym} &  &  & \checkmark &  & 2017-11-05 \\
    \href{https://cran.r-project.org/web/packages/pgsc/index.html}{\texttt{pgsc}} &\texttt{R} & \texttt{SC} & \texttt{Perm} & \checkmark &  &  &  & 2018-10-28 \\
    \href{https://cran.r-project.org/web/packages/MSCMT/index.html}{\texttt{MSCMT}} & \texttt{R} & \texttt{SC} & \texttt{Perm} & &  &  & \checkmark &  2019-11-14 \\
    \href{https://ideas.repec.org/c/boc/bocode/s458398.html}{\texttt{npsynth}} & \texttt{St} & \texttt{SC} & \texttt{Perm} & & & & & 2020-06-23\\
    \href{https://cran.r-project.org/web/packages/tidysynth/index.html}{\texttt{tidysynth}} & \texttt{R} & \texttt{SC} & \texttt{Perm}&  &  &  &   & 2021-01-27 \\ \multicolumn{9}{c}{\hdashrule[0.5ex]{22.5cm}{1pt}{1pt}} \\ 
    \href{https://cran.r-project.org/web/packages/microsynth/index.html}{\texttt{microsynth}}  & \texttt{R} & \texttt{CA} & \texttt{Perm}& \checkmark &  &  & \checkmark & 2021-02-26 \\
    \href{https://github.com/kwuthrich/scinference}{\texttt{scinference}} & \texttt{R} & \texttt{SC}, \texttt{LA}& \texttt{Perm} & &  & \checkmark &   &  2021-05-14 \\
    \href{https://github.com/hollina/scul}{\texttt{SCUL}} &  \texttt{R} & \texttt{LA} & \texttt{Perm}&  &  &  &   & 2021-05-19 \\
    \href{https://cran.r-project.org/web/packages/gsynth/index.html}{\texttt{gsynth}} &  \texttt{R} & \texttt{FA} & \texttt{Asym} & \checkmark & \checkmark &  & \checkmark & 2021-08-06 \\
    \href{https://pypi.org/project/SyntheticControlMethods}{\texttt{Synth}} & \texttt{Py} & \texttt{SC} & \texttt{Perm} &  &  &  &   & 2021-10-07 \\    
    \multicolumn{9}{c}{\hdashrule[0.5ex]{22.5cm}{1pt}{1pt}} \\
    \href{https://pypi.org/project/treebased-synthetic-controls}{\texttt{treebased-sc}} &  \texttt{Py} & \texttt{TB} & \texttt{Perm} &  &  & \checkmark &  & 2021-11-01 \\
    \href{https://cran.r-project.org/web/packages/SynthCast/index.html}{\texttt{SynthCast}} &  \texttt{R} & \texttt{SC} & \texttt{Perm}&  &  &  &   & 2022-03-08 \\
    \href{https://github.com/synth-inference/synthdid}{\texttt{sytnhdid}} & \texttt{R} & \texttt{LS}, \texttt{RI} & \texttt{Asym} & \checkmark &\checkmark &  &  & 2022-03-15\\    
    \href{https://ideas.repec.org/c/boc/bocode/s459076.html}{\texttt{allsynth}} & \texttt{St} & \texttt{SC} & \texttt{Perm} & \checkmark & \checkmark & & & 2022-05-07\\    
    \href{https://ideas.repec.org/c/boc/bocode/s459017.html}{\texttt{synth2}} & \texttt{St} & \texttt{SC} & \texttt{Perm} & & & & & 2022-05-28\\  
    \multicolumn{9}{c}{\hdashrule[0.5ex]{22.5cm}{1pt}{1pt}} \\
    \href{https://cran.r-project.org/web/packages/Synth/index.html}{\texttt{Synth}} & \texttt{R}, \texttt{St} & \texttt{SC} & \texttt{Perm} &  &  &  &   & 2022-06-08 \\
    \href{https://cran.r-project.org/web/packages/SCtools/index.html}{\texttt{SCtools}} & \texttt{R} & \texttt{SC} & \texttt{Perm}& \checkmark &  &  & \checkmark  & 	2022-06-09 \\
    \href{https://github.com/ebenmichael/augsynth}{\texttt{augsynth}} & \texttt{R} & \texttt{SC}, \texttt{RI} & \texttt{Perm} & \checkmark & \checkmark &  &  & 2022-08-02 \\
    \href{https://ideas.repec.org/c/boc/bocode/s459107.html}{\texttt{scul}} & \texttt{St} & \texttt{LA} & \texttt{Perm} & & & & & 2022-08-21\\    
    \midrule
    \href{https://nppackages.github.io/scpi/}{\texttt{scpi}}  & \texttt{Py}, \texttt{R}, \texttt{St} & \texttt{SC}, \texttt{LA}, \texttt{RI}, \texttt{LS}, \texttt{+} & \texttt{PI}, \texttt{Asym}, \texttt{Perm} & \checkmark & \checkmark & \checkmark & \checkmark & 2022-10-07\\
    \midrule
    \bottomrule
    \end{tabular}%
  }
  \label{tab:comparison}%
  \par
 	\begin{center}
		\parbox[1]{\textwidth}{\footnotesize \textit{Note:} \texttt{Py} = \texttt{Python} (\url{https://www.python.org/}); \texttt{R} = \texttt{R} (\url{https://cran.r-project.org/}); \texttt{St} = \texttt{Stata} (\url{https://www.stata.com/}); \texttt{LA} = Lasso penalty; \texttt{CA} = calibration; \texttt{FA} = factor-augmented models; \texttt{LS} = unconstrained least squares; \texttt{RI} = Ridge penalty; \texttt{SC} = canonical synthetic control; \texttt{TB} = tree-based methods; \texttt{+} = user-specified options (see Table \ref{tab:lit_constr} below for more details); \texttt{Perm} = permutation-based inference; \texttt{Asym} = asymptotic-based inference; \texttt{PI} = prediction intervals (non-asymptotic probability guarantees). The symbol {\checkmark} means that the feature is available. The last column reports the date of last update as of October 7, 2022.}
	\end{center}  
\end{table}

The rest of the article is organized as follows. Section \ref{sec: setup} introduces the canonical synthetic control setup, and also briefly discusses extensions to multiple treated units with possibly staggered treatment adoption. Section \ref{sec: point prediction} gives a brief introduction to the theory and methodology underlying the point estimation/prediction for synthetic control methods, discussing implementation details. Section \ref{sec: uncertainty quantification} gives a brief introduction to the theory and methodology underlying the uncertainty quantification via prediction intervals for synthetic control methods, and also discusses the corresponding issues of implementation. Section \ref{sec: illustration} showcases some of the functionalities of the package using a real-world dataset, and Section \ref{sec:conclusion} concludes. The appendices illustrate the \texttt{Python} (Appendix \ref{app:python}) and \texttt{Stata} (Appendix \ref{app:stata}) implementations of \texttt{scpi}. Detailed instructions for installation, script files to replicate the analyses, links to software repositories, and other companion information can be found in the package's website, \url{https://nppackages.github.io/scpi/}.

\section{Setup}\label{sec: setup}

We first consider the canonical synthetic control framework with a single treated unit. The researcher observes $J+1$ units for $T_0+T_1$ periods of time. Units are indexed by $i = 1,2,\ldots J, J+1$, and time periods are indexed by $t=1,2, \ldots,T_0, T_0+1, \ldots, T_0+T_1$. During the first $T_0$ periods, all units are untreated. Starting at $T_0 + 1$, unit $1$ receives treatment but the other units remain untreated. Once the treatment is assigned at $T_0+1$, there is no change in treatment status: the treated unit continues to be treated  and the untreated units remain untreated until the end of the series, $T_1$ periods later. The single treated unit in our context could be understood as an ``aggregate'' of multiple treated units; see Section \ref{sec: extension} below for more discussion. 

Each unit $i$ at period $t$ has two potential outcomes, $Y_{it}(1)$ and $Y_{it}(0)$, respectively denoting the outcome under treatment and the outcome in the absence of treatment. Two implicit assumptions are imposed: no spillovers (the potential outcomes of unit $i$ depend only on $i$'s treatment status) and no anticipation (the potential outcomes at $t$ depend only on the treatment status of the same period). Then, the observed outcome $Y_{it}$ is
    \[Y_{it}=\begin{cases} 
    Y_{it}(0),\quad & \text{if} \quad i \in \{2,\ldots, J+1\} \\
    Y_{it}(0),\quad & \text{if} \quad i=1 \text{ and } t \in \{1,\ldots,T_0\} \\ 
    Y_{it}(1),\quad & \text{if} \quad i=1 \text{ and } t \in \{T_0+1,\ldots,T_0+T_1\}\end{cases}.
    \]

The causal quantity of interest is the difference between the outcome path taken by the treated unit, and the path it would have taken in the absence of the treatment:
\[
\tau_t:=Y_{1t}(1)-Y_{1t}(0), \qquad t>T_0.
\]
We view the two potential outcomes $Y_{1t}(1)$ and $Y_{1t}(0)$ as random variables, which implies that $\tau_t$ is a random quantity as well, corresponding to the treatment effect on a \textit{single} treated unit. This contrasts with other analysis that regards the treatment effect as a fixed parameter \citep[see][for references]{Abadie_2021_JEL}.

The potential outcome $Y_{1t}(1)$ of the treated unit is observed after the treatment. To recover the treatment effect $\tau_t$, it is necessary to have a ``good'' prediction of the counterfactual outcome $Y_{1t}(0)$ of the treated after the intervention. The idea of the synthetic control method is to find a vector of weights $\bw = (w_2,w_3,\dots,w_{J+1})'$ such that a given loss function is minimized under constraints, only using pre-intervention observations. Given the resulting set of constructed weights $\widehat{\bw}$, the treated unit's counterfactual (potential) outcome is calculated as $\widehat{Y}_{1t}(0) = \sum_{i=2}^{J+1} \widehat{w}_{i}Y_{it}(0)$ for $t>T_0$. The weighted average $\widehat{Y}_{1t}(0)$ is often referred to as the \textit{synthetic control} of the treated unit, as it represents how the untreated units can be combined to provide the best counterfactual for the treated unit in the post-treatment period. In what follows, we briefly describe different approaches for point estimation/prediction leading to $\widehat{Y}_{1t}(0)$, and then summarize the uncertainty quantification methods to complement those predictions.

\subsection{Extensions} \label{sec: extension}

Building on the canonical synthetic control setup, we can consider other settings involving multiple treated units with possibly staggered treatment adoption. In particular, we briefly discuss three potential extensions of practical interest. 

\begin{itemize}[leftmargin=*]

    \item \textbf{Multiple post-treatment periods}. When outcomes are observed in multiple periods after the treatment, a researcher might be interested in the average treatment effect on the (single) treated unit across  multiple post-treatment periods rather than the effect at a single period:
    \[
    \tau:= \frac{1}{T_1}\sum_{t=T_0+1}^{T_0+T_1}\Big(Y_{1t}(1)-Y_{1t}(0)\Big)=\frac{1}{T_1}\sum_{t=T_0+1}^{T_0+T_1}\tau_t.
    \]
    The analysis of this quantity can be accommodated by the framework above. For instance, given the predicted counterfactual outcome $\widehat{Y}_{1t}(0) = \sum_{i=2}^{J+1} \widehat{w}_{i}Y_{it}(0)$ for each post-treatment period $t>T_0$, the predicted average counterfactual outcome of the treated is given by 
    \[
    \sum_{i=2}^{J+1} \widehat{w}_{i}
    \Big(\frac{1}{T_1}\sum_{t=T_0+1}^{T_0+T_1}Y_{it}(0)\Big).
    \] 
    This construction is equivalent to regarding the $T_1$ post-treatment periods as a ``single'' period and defining the post-treatment predictors as averages of the corresponding predictors across post-treatment time periods.

    \item \textbf{Multiple treated units}. The canonical single treated unit framework above can also be extended to the more general case of multiple treated units. For instance, suppose a researcher observes $N_0+N_1$ units for $T_0+T_1$ time periods, and let units be indexed by $i=1, \ldots, N_1, N_1+1, \ldots,N_0 + N_1$. Without loss of generality, the first $1$ to $N_1$ units are assumed to be treated and units from $N_1+1$ to $N_0$ to be untreated. Treated and untreated potential outcomes are, respectively, denoted by $Y_{it}(1)$ and $Y_{it}(0)$ for $i=1, \ldots,N_0 + N_1$. The observed outcome of the $i$th treated unit is given by ${Y_{it}:=\I(t\leq T_0)Y_{it}(0)+\I(t>T_0)Y_{it}(1)}$. 
    
    In such setting, a researcher might be interested in the \textit{individual} treatment effect $\tau_{it}$ 
    \[\tau_{it} := Y_{it}(1)-Y_{it}(0), \quad t>T_0, \quad i=1, \ldots, N_1,\]
    or in the \textit{average} treatment effect on the treated $\tau_{\cdot t}$ across treated units
    \[\tau_{\cdot t}:=\frac{1}{N_1}\sum_{j=1}^{N_1}\Big(Y_{jt}(1)-Y_{jt}(0)\Big), \qquad t>T_0.\]
    The first causal quantity, $\tau_{it}$, can be predicted in the framework described above considering one treated unit \textit{at a time} or, alternatively, by considering all $N_1$ treated units \textit{jointly}.
    
    To predict the second causal quantity, $\tau_{\cdot t}$, one extra step is necessary. Define an aggregate unit ``\texttt{ave}'' whose observed outcome is $Y_{t}^{\texttt{ave}}:=\frac{1}{N_1}\sum_{j=1}^{N_1}Y_{jt}$, for $t=1, \ldots, T_0+T_1$. Other features of ``unit 1'' used in the synthetic control construction can be defined similarly as averages of the corresponding features across multiple treated units. The framework above can now be applied to the ``new'' average unit with outcome $Y_{t}^{\texttt{ave}}$.

    \item \textbf{Staggered treatment adoption}.
    Our framework can also be extended to the scenario where multiple treated units are assigned to treatment at different points in time, a \textit{staggered adoption} design. In this case, one can understand the adoption time as a multivalued treatment assignment, and a large class of causal quantities can be defined accordingly. For example, let $T_i\in\{T_0+1, T_0+2, \ldots, T, \infty\}$ denote the adoption time of unit $i$ where $T_i=\infty$ means unit $i$ is never treated, and $Y_{it}(s)$ represents the potential outcome of unit $i$ at time $t$ that would be observed if unit $i$ had adopted the treatment at time $s$. Suppose that the treatment effect on unit $i$ one period after the treatment, i.e., $Y_{i(T_i+1)}(T_i)-Y_{i(T_i+1)}(\infty)$, is of interest. One can take all units that are treated later than $T_i+1$ to obtain the synthetic control weights and construct the synthetic control prediction of the counterfactual outcome $Y_{i(T_i+1)}(\infty)$ accordingly. The methodology described below can be immediately applied to this problem.
\end{itemize}

The package \texttt{scpi} allows for estimation/prediction of treatment effects and uncertainty quantification via prediction intervals for the more general synthetic control settings discussed above. However, in order to streamline the exposition, the rest of this article focuses on the case of a single treated unit. See \cite*{Cattaneo-Feng-Palomba-Titiunik_2022_wp} for a formal treatment of more general staggered adoption problems, and its supplemental appendix for further details on how the package \texttt{scpi} can be used in settings with multiple treatment units and staggered treatment adoption. Our companion replication files illustrate both the canonical single treated unit framework and the generalizations discussed above.

\section{Synthetic Control Prediction}\label{sec: point prediction}

We consider synthetic control weights constructed simultaneously for $M$ features of the treated unit, denoted by $\bA_l=(a_{1,l}, \cdots, a_{T_0,l})'\in\mathbb{R}^{T_0}$, with index $l=1,\ldots, M$. For each feature $l$, there exist $J+K$ variables that can be used to predict or ``match'' the $T_0$-dimensional vector $\bA_l$. These $J+K$ variables are separated into two groups denoted by $\bB_l=(\bB_{1,l}, \bB_{2,l}, \cdots, \bB_{J, l})\in\mathbb{R}^{T_0\times J}$ and $\bC_l=(\bC_{1,l}, \cdots, \bC_{K,l})\in\mathbb{R}^{T_0\times K}$, respectively. More precisely, for each $j$, $\bB_{j,l}=(b_{j1,l}, \cdots, b_{jT_0,l})'$ corresponds to the $l$th feature of the $j$th unit observed in $T_0$ pre-treatment periods and, for each $k$, $\bC_{k,l}=(c_{k1,l}, \cdots, c_{kT_0,l})'$ is another vector of control variables also possibly used to predict $\bA_l$ over the same pre-intervention time span. For ease of notation, we let $d=J+KM$. 

The goal of the synthetic control method is to search for a vector of common weights $\bw\in\mathcal{W}\subseteq\mathbb{R}^{J}$ across the $M$ features and a vector of coefficients $\br\in\mathcal{R}\subseteq\mathbb{R}^{KM}$, such that the linear combination of $\bB_l$ and $\bC_l$ ``matches'' $\bA_l$ as close as possible, during the pre-intervention period, for all $1\leq l\leq M$ and some convex feasibility sets $\mathcal{W}$ and $\mathcal{R}$ that capture the restrictions imposed. Specifically, we consider the following optimization problem:
\begin{equation}\label{eq: estimated weight}
\widehat{\bbeta} := (\widehat{\bw}',\;\widehat{\br}')' \in\underset{\bw\in\mathcal{W},\, \br\in\mathcal{R}}{\argmin}\;
(\bA-\bB\bw-\bC\br)'\bV(\bA-\bB\bw-\bC\br)
\end{equation}
where
\[\underbrace{\mathbf{A}}_{T_0\cdot M\times 1} = \begin{bmatrix} \mathbf{A}_1 \\ \vdots \\ \mathbf{A}_M \end{bmatrix}, \quad \underbrace{\mathbf{B}}_{T_0\cdot M \times J} = \begin{bmatrix} \mathbf{B}_1 \\ \vdots \\ \mathbf{B}_M \end{bmatrix},\quad \underbrace{\mathbf{C}}_{T_0\cdot M \times K\cdot M}=\begin{bmatrix} \mathbf{C}_1 & \mathbf{0} & \cdots & \mathbf{0} \\
\mathbf{0} & \mathbf{C}_2 & \cdots & \mathbf{0} \\
\vdots & \vdots &\ddots &\vdots \\
\mathbf{0} & \mathbf{0} & \cdots & \mathbf{C}_M\end{bmatrix}\]
and $\bV$ is a $T_0\cdot M\times T_0\cdot M$ weighting matrix reflecting the relative importance of different equations and time periods.

From \eqref{eq: estimated weight}, we can define the pseudo-true residual $\bu$ as
\begin{equation}\label{eq: pseudo true value}
\bu=\bA-\bB\bw_0-\bC\br_0,
\end{equation}
where $\bw_0$ and $\br_0$ denote the mean squared error population analog of $\widehat{\bw}$ and $\widehat{\br}$. As discussed in the next section, the proposed prediction intervals are valid conditional on some information set $\mathscr{H}$. Thus, $\bw_0$ and $\br_0$ above are viewed as the (possibly constrained) best linear prediction coefficients conditional on $\mathscr{H}$. We \textit{do not} attach any structural meaning to $\bw_0$ and $\br_0$: they are only (conditional) pseudo-true values whose meaning should be understood in context, and are determined by the assumptions imposed on the data generating process. In other words, we allow for misspecification when constructing the synthetic control weights $\widehat{\bw}$, as this is the most likely scenario in practice.

Given the constructed weights $\widehat{\bw}$ and coefficients $\widehat{\br}$, the  counterfactual outcome at the post-treatment period $T$ for the treated unit, $Y_{1T}(0)$, is predicted by
\begin{equation}\label{eq: SC prediction}
\widehat{Y}_{1T}(0) = \bx_T'\widehat{\bw}+\bg_T'\widehat{\br} = \bp_T'\widehat{\bbeta}, \qquad \bp_T := (\bx_T', \,\bg_T')', \qquad T>T_0,
\end{equation}
where $\bx_T\in\mathbb{R}^{J}$ is a vector of predictors for control units observed in time $T$ and $\bg_T\in\mathbb{R}^{KM}$ is another set of user-specified predictors observed at time $T$. Variables included in $\bx_T$ and $\bg_T$ need not be the same as those in $\bB$ and $\bC$, but in practice it is often the case that $\bx_T=(Y_{2T}(0), \cdots, Y_{(J+1)T}(0))'$ and $\bg_T$ is excluded when $\bC$ is not specified.

The next section discusses implementation details leading to $\widehat{Y}_{1T}(0)$, including the choice of feasibility sets $\mathcal{W}$ and $\mathcal{R}$, weighting matrix $\bV$, and additional covariates $\bC$.

\subsection{Implementation}\label{sec: sc implementation}

The function \texttt{scdata()} in \texttt{scpi} prepares the data for point estimation/prediction purposes. This function takes as input an object of class \texttt{DataFrame} and outputs an object of class \texttt{scpi\_data} containing the matrices $\bA,\bB,\bC$ described above, and a matrix of post-treatment predictors $\bP=(\bp_{T_0+1}, 
\cdots, \bp_{T_0+T_1})'$. The user must provide a variable containing a unit identifier (\texttt{id.var}), a time variable (\texttt{time.var}), an outcome variable (\texttt{outcome.var}), the features to be matched (\texttt{features}), the treated unit (\texttt{unit.tr}), the control units (\texttt{unit.co}), the pre-treatment periods (\texttt{period.pre}), and the post-treatment periods (\texttt{period.post}). These options completely specify $\bA,\bB$, and $\bP$.  The user can also control the form of $\mathcal{R}$ in \eqref{eq: estimated weight} or, equivalently, the form of $\bC$, through the options \texttt{cov.adj} and \texttt{constant}. The former option allows the user to flexibly specify covariate adjustment feature by feature, while the latter option introduces a column vector of ones of size $T_0\cdot M$ in $\bC$. If $M=1$, this is a simple constant term, but if $M\geq 2$ it corresponds to an intercept which is common across features. 

The use of the options \texttt{cov.adj} and \texttt{constant} is best explained through some examples. If the user specifies only one feature ($M=1$), then \texttt{cov.adj} can be an unnamed list:

\begin{lstlisting}
cov.adj <- list(c("constant","trend"))
\end{lstlisting}
This particular choice includes a constant term and a linear time trend in $\bC$. If instead multiple features ($M \geq 2$) are used to find the synthetic control weights $\widehat\bw$, then \texttt{cov.adj} allows for feature-specific covariate adjustment. For example, in a two-feature setting ($M=2$), the code

\begin{lstlisting}
cov.adj <- list('f1' = c("constant","trend"),'f2' = c("trend"))
\end{lstlisting}
specifies $\bC$ as a block diagonal matrix where the first block $\bC_1$ contains a constant term and a trend, while the second block $\bC_2$ contains only a trend. If the user wants all features to share the same covariate adjustment, then it is sufficient to input a list with a unique element:

\begin{lstlisting}
cov.adj <- list(c("constant","trend"))
\end{lstlisting}

This specification creates a block diagonal matrix $\bC$ with identical blocks. In the same example with $M=2$, if \texttt{constant <- TRUE} and \texttt{cov.adj <- NULL}, then $\bC$ would not be block diagonal, but rather a column vector of ones of size $2 T_0$.

Finally, if $\bA$ and $\bB$ form a cointegrated system, by setting the option \texttt{cointegrated.data} to \texttt{TRUE} in \texttt{scdata()}, the matrix $\bP$ is prepared in such a way that the function \texttt{scpi()} will properly handle in-sample and out-of-sample uncertainty quantification (see sections \ref{sec: how to in sample} and \ref{sec:how to out of sample}).

Once all the design matrices $\bA,\bB,\bC,$ and $\bP$ have been created, we can proceed with point estimation/prediction of the counterfactual outcome of interest via the function \texttt{scest()}.

The form of the feasibility set $\mathcal{W}$ in \eqref{eq: estimated weight} or, equivalently, the constraints imposed on the weights $\bw$, can be set using the option \texttt{w.constr}. The package allows for the following family of constraints:
\begin{align*}
    \mathcal{W} \in \Big\{ \mathbb{R}^J, \;\{\bw \in \mathbb{W}: ||\bw||_p \leq Q\}, \;
    \{\bw \in \mathbb{R}^J: ||\bw||_1 = Q, ||\bw||_2 \leq Q_2\} \Big\}, \\ \mathbb{W}\in\{\mathbb{R}^J,\mathbb{R}^J_+\}, \qquad p \in \{1,2\}, \qquad Q \in \mathbb{R}_{++}, \qquad Q_2 \in \mathbb{R}_{++},\qquad 
\end{align*}
where the inequality constraint on the norm can be made an equality constraint as well. The user can specify the desired form for $\mathcal{W}$ through a list to be passed to the option \texttt{w.constr}:
\begin{lstlisting}
W1 <- list(p = "no norm", lb = -Inf)
W2 <- list(p = "L1", dir = "==", Q = 1, lb = 0)
W3 <- list(p = "L2", dir = "<=", Q = 1, lb = -Inf)
W4 <- list(p = "L1-L2", lb = -Inf, Q = 1, Q2 = 1, dir = "==/<=")
\end{lstlisting}
The four lines above create $\mathcal{W}_1 = \mathbb{R}^J$, $\mathcal{W}_2 = \{\bw \in \mathbb{R}_+^{J}: ||\bw||_1= 1\}$, $\mathcal{W}_3 = \{\bw \in \mathbb{R}^{J}: ||\bw||_2\leq 1\}$, and $\mathcal{W}_4 = \{\bw \in \mathbb{R}^{J}: ||\bw||_1= 1, ||\bw||_2\leq 1\}$, respectively. In greater detail,
\begin{itemize}
    \item\texttt{p} chooses the constrained norm of $\bw$ among the options \texttt{`no norm'}, \texttt{`L1'}, \texttt{`L2'}, or \texttt{`L1-L2'}
    \item \texttt{dir} sets the direction of the constraint $\|\bw\|_p$ and it can be either \texttt{`=='}, \texttt{`<='}, or \texttt{`==/<='}
    \item \texttt{Q} is the size of the constraint and it can be set to any positive real number
    \item \texttt{lb} sets a  (common) lower bound on $\bw$ and it takes as input either \texttt{0} or \texttt{-Inf}
\end{itemize}

Popular constraints can be called explicitly using the option \texttt{name} in the list passed to \texttt{w.constr}. Table \ref{tab:pop_constr} gives prototypical examples of such constraints.
\begin{table}[H]
  \centering
  \caption{Constraints on the weights that can be directly called.}
    \begin{tabular}{ccc}
    \toprule\midrule
       Name   & \multicolumn{1}{c}{\texttt{w.constr}} & $\mathcal{W}$ \\
    \midrule
     OLS   & \texttt{list(name = `ols')} & $\mathbb{R}^J$  \\
   simplex & \texttt{list(name = `simplex', Q = Q)} &  $\{\bw \in \mathbb{R}_+^J: ||\bw ||_1 = Q\}$\\
    lasso & \texttt{list(name = `lasso', Q = Q)} & $\{\bw \in \mathbb{R}^J: ||\bw ||_1 \leq Q\}$  \\
    ridge & \texttt{list(name = `ridge', Q = Q)} & $\{\bw \in \mathbb{R}^J: ||\bw ||_2 \leq Q\}$ \\
    L1-L2 & \texttt{list(name = `L1-L2', Q = Q, Q2 = Q2)} & $\{\bw \in \mathbb{R}_+^J:||\bw ||_1 = Q, ||\bw ||_2 \leq Q_2\}$ \\
    \midrule\bottomrule
    \end{tabular}%
  \label{tab:pop_constr}%
\end{table}%
In particular, specifying \texttt{list(name = `simplex', Q = 1)} gives the standard constraint used in the canonical synthetic control method, that is, computing weights in \eqref{eq: estimated weight} such that they are non-negative and sum up to one, and without including an intercept. This is the default in the function \texttt{scest()} (and \texttt{scpi()}). The following snippet showcases how each of these five constraints can be called automatically through the option \texttt{name} and manually through the options \texttt{p}, \texttt{Q}, \texttt{Q2}, \texttt{lb}, and \texttt{dir}. In the snippet, \texttt{Q} and \texttt{Q2} are set to 1 for ridge and L1-L2 constraints, respectively, for simplicity, but to replicate the results obtained with the option \texttt{name} one should input the proper \texttt{Q} according to the rules of thumb described further below.

{\singlespacing
\begin{lstlisting}
## Simplex
w.constr <- list(name = "simplex")
w.constr <- list(p = "L1", lb = 0, Q = 1, dir = "==")

## Least Squares
w.constr <- list(name = "ols")
w.constr <- list(p = "no norm", lb = -Inf, Q = NULL, dir = NULL)

## Lasso
w.constr <- list(name = "lasso")
w.constr <- list(p = "L1", lb = -Inf, Q = 1, dir = "<=")

## Ridge
w.constr <- list(name = "ridge")
w.constr <- list(p = "L2", lb = -Inf, Q = 1, dir = "<=")

## L1-L2
w.constr <- list(name = "L1-L2")
w.constr <- list(p = "L1-L2", lb = 0, Q = 1, Q2 = 1, dir = "==/<=")
\end{lstlisting}}

Using the option \texttt{w.constr} in \texttt{scest()} (or \texttt{scpi()}) and the options \texttt{cov.adj} and \texttt{constant} in \texttt{scdata()} appropriately, i.e., setting $\mathcal{W}$ and $\mathcal{R}$ in \eqref{eq: estimated weight}, many synthetic control estimators proposed in the literature can be implemented. Table \ref{tab:lit_constr} provides a non-exhaustive list of such examples. 

\begin{table}[H]
  \centering
  \renewcommand{\arraystretch}{1.1}
  \caption{Examples of $\mathcal{W}$ and $\mathcal{R}$ in the synthetic control literature ($M=1$).}
  \resizebox{\textwidth}{!}{
    \begin{tabular}{lcccccc}
    \toprule\midrule
    \multirow{2}[0]{*}{Article} & \multirow{2}[0]{*}{$\mathcal{W}$} & \multirow{2}[0]{*}{$\mathcal{R}$} & \multicolumn{3}{c}{\texttt{w.constr}} & \multirow{2}[0]{*}{\texttt{constant}} \\ \cline{4-6}
          &       &       & \texttt{name}  & \texttt{Q} & \texttt{Q2}  &  \\
   \midrule
    \cite{Hsiao-et-al_2012_JAE} & $\mathbb{R}^J$ & $\mathbb{R}$ & \texttt{"ols"} & \texttt{NULL} & \texttt{NULL} & \texttt{TRUE} \\
    \cite{Abadie-Diamond-Hainmueller_2010_JASA} & $\{\bw\in \mathbb{R}^J_+: ||\bw||_1 = 1\}$ & $\{0\}$ & \texttt{"simplex"} & \texttt{1} & \texttt{NULL} & \texttt{FALSE} \\
    \cite{Ferman-Pinto_2021_QE} & $\{\bw\in\mathbb{R}^J_+:||\bw||_1 = 1\}$ & $\mathbb{R}$ & \texttt{"simplex"} &  \texttt{1} & \texttt{NULL} & \texttt{TRUE}\\
    \cite{Chernozhukov-Wuthrich-Zhu_2021_JASA} & $\{\bw \in \mathbb{R}^J:||\bw||_1 \leq 1\}$ & $\mathbb{R}$ & \texttt{"lasso"}& \texttt{1} & \texttt{NULL} & \texttt{TRUE}\\
    \cite{Amjad-Shah-Shen_2018_JMLR} & $\{\bw\in \mathbb{R}^J: ||\bw||_2 \leq Q\}$ & $\{0\}$ & \texttt{"ridge"} & \texttt{Q} & \texttt{NULL}  & \texttt{FALSE}\\
    \cite{Arkhangelsky-et-al_2021_AER} & $\{\bw\in \mathbb{R}_+^J: ||\bw||_1 =1, ||\bw||_2 \leq Q_2\}$ & $\mathbb{R}$ & \texttt{"L1-L2"} & 1 & \texttt{Q}  & \texttt{TRUE}\\
    \midrule\bottomrule
    \end{tabular}}
  \label{tab:lit_constr}%
\end{table}

\subsubsection*{Tuning parameter choices}

We provide rule-of-thumb choices of the tuning parameter $Q$ for Lasso- and Ridge-type constraints.
\begin{itemize}[leftmargin=*]
    \item \textbf{Lasso $(p=1)$.} Being Lasso similar in spirit to the ``simplex''-type traditional constraint in the synthetic control literature, we propose $Q=1$ as a rule of thumb.
    
    \item \textbf{Ridge $(p=2)$.} 
    It is well known that the Ridge prediction problem can be equivalently formulated as an unconstrained penalized optimization problem and as a constrained optimization problem.
    More precisely, assuming $\bC$ is not used and $M=1$ for simplicity, the two Ridge-type problems are
    \[\widehat \bw := \argmin_{\bw \in \mathbb{R}^J} (\bA - \bB\bw)^\prime \bV (\bA -\bB\bw) + \lambda ||\bw||_2^2,\]
    where $\lambda\geq 0$ is a shrinkage parameter, and
    \[\widehat \bw := \argmin_{\bw \in \mathbb{R}^J,\, ||\bw||_2^2 \leq Q^2} (\bA - \bB\bw)^\prime \bV (\bA -\bB\bw),\]
    where $Q\geq 0$ is the (explicit) size of the constraint on the norm of $\bw$.
    Under the assumption of Gaussian errors, a risk-minimizing choice \citep*{Hoerl-Kannard-Kent_1975_ridge} of the standard shrinkage tuning parameter is
    \[\lambda= J \widehat\sigma_{\mathtt{OLS}}^{2}/\|\widehat\bw_{\mathtt{OLS}}\|_{2}^{2},\]
    where $\widehat{\sigma}_{\mathtt{OLS}}^2$ and $\widehat\bw_{\mathtt{OLS}}$ are estimators of the variance of the pseudo-true residual $\bu$ and the coefficients $\bw_0$ based on least squares regression, respectively.
    
    Since the two optimization problems above are equivalent, there exists a one-to-one correspondence between $\lambda$ and $Q$. For example, assuming the columns of $\bB$ are orthonormal, the closed-form solution for the Ridge estimator is $\widehat\bw = (\bI + \lambda \bI)^{-1}\widehat\bw_{\mathtt{OLS}}$, and it follows that if the constraint on the $\ell^2$-norm is binding, then $Q = ||\widehat\bw||_2 = ||\widehat\bw_{\mathtt{OLS}}||_2/(1+\lambda)$. \smallskip
    
    However, if $J > T_0$, $\widehat\bw_{\mathtt{OLS}}$ does not exist, hence we cannot rely on the approach suggested above. Indeed, the proposed mapping between $\lambda$ and $Q$ is ill-defined and, also, we are unable to estimate $\lambda$. In this case, we first make the design low-dimensional by performing variable selection on $\bB$ with Lasso. Once we select the columns of $\bB$ whose Lasso coefficient is non-zero, we choose $\lambda$ according to the rule of thumb described above.
    
    If more than one feature is specified ($M>1$),  we compute the size of the constraint $Q_l$ for each feature $l = 1,\ldots, M$ and then select $Q$ as the tightest constraint to favor shrinkage of $\bw$, that is $Q:=\min_{l=1,\ldots,M}Q_l$. 
\end{itemize}

\subsubsection*{Missing Data} 

In case of missing values, we adopt different strategies depending on which units have missing entries and when these occur.
\begin{itemize}
    \item \textit{Missing pre-treatment data}. In this case we compute $\widehat \bw$ without the periods for which there is at least a missing entry for either the treated unit or one of the donors.
    
    \item \textit{Missing post-treatment donor data.} Suppose that the $i$th donor has a missing entry in one of the $M$ features in the post-treatment period $\widetilde T$. It implies that the predictor vector $\bp_{\widetilde T}$ has a missing entry, and thus the 
    synthetic unit and the associated prediction intervals are not available.
    
    \item \textit{Missing post-treatment treated data.} Data for the treated unit after the treatment is only used to quantify the treatment effect $\tau_T$, which then will not be available. However, prediction intervals for the synthetic point prediction of the counterfactual outcome $Y_{1t}(0)$, $t>T_0$, can still be computed in the usual way as they do not rely on the availability of such data points.
\end{itemize}

\section{Uncertainty Quantification}\label{sec: uncertainty quantification}
Following \citet*{Cattaneo-Feng-Titiunik_2021_JASA} and \citet*{Cattaneo-Feng-Palomba-Titiunik_2022_wp}, we view the quantity of interest $\tau_{T}$ within the synthetic control framework as a random variable, and hence we refrain from calling it a parameter. Building an analogy with the concept of estimand (or parameter of interest), we refer to $\tau_T$ as a \textit{predictand}. Consequently, we prefer to call $\widehat{\tau}_{T}=Y_{1T}(1)-\widehat{Y}_{1T}(0)$ based on \eqref{eq: SC prediction} a \textit{prediction} of $\tau_{T}$ rather than an \textit{estimator} of it, and our goal is to characterize the uncertainty of $\widehat{\tau}_T$ by building prediction intervals rather than confidence intervals. In practice, it is appealing to construct prediction intervals that are valid \textit{conditional} on a set of observables. We let $\mathscr{H}$ be an information set generated by all features of control units and covariates used in the synthetic control construction, i.e., $\bB$, $\bC$, $\bx_T$, and $\bg_T$.

We first decompose the potential outcome of the treated unit based on $\bw_0$ and $\br_0$ introduced in \eqref{eq: pseudo true value}:
\begin{equation}\label{eq: SC Y1T(0)}
    Y_{1T}(0) \equiv \bx_T'\bw_0+\bg_T'\br_0+e_T = \bp_T'\bbeta_0 + e_T, \qquad T>T_0,
\end{equation}
where $e_T$ is defined by construction. In our analysis, $\bw_0$ and $\br_0$ are assumed to be (possibly) random quantities around which $\widehat{\bw}$ and $\widehat{\br}$ are concentrating in probability, respectively. Then, the distance between the predicted treatment effect on the treated and the target population one is
\begin{equation}\label{eq: tauhat - tau}
    \widehat{\tau}_{T} - \tau_{T} = Y_{1T}(0) - \widehat{Y}_{1T}(0) = e_T-\bp_T'(\widehat\bbeta-\bbeta_0).
\end{equation}
where $e_T$ is the out-of-sample error coming from misspecification along with any additional noise occurring at the post-treatment period $T > T_0$, and the term $\bp_T'(\widehat\bbeta-\bbeta_0)$ is the in-sample error coming from the construction of the synthetic control weights. Our goal is to find probability bounds on the two terms separately to give uncertainty quantification: 
for some pre-specified levels $\alpha_1, \alpha_2\in(0,1)$, with high probability over $\mathscr{H}$, 
\begin{align*}
 \P\big[M_{1,\mathtt{L}}\leq\bp_T'(\widehat{\bbeta}-\bbeta_0)\leq M_{1,\mathtt{U}} \; \big| \; \mathscr{H} \big]\geq 1-\alpha_1 
 \quad \text{and} \quad
 \P\big[M_{2,\mathtt{L}}\leq e_T\leq M_{2,\mathtt{U}} \; \big| \; \mathscr{H} \big] \geq 1-\alpha_2.
\end{align*}
It follows that these probability bounds can be combined to construct a prediction interval for $\tau_T$ with conditional coverage at least $1-\alpha_1-\alpha_2$: with high probability over $\mathscr{H}$, 
\[
\P\big[\widehat{\tau}_T+M_{1,\mathtt{L}}-M_{2,\mathtt{U}} \leq \tau_T\leq 
\widehat{\tau}_T+M_{1,\mathtt{U}}-M_{2,\mathtt{L}} \big|  \mathscr{H}  \big]
	\geq 1-\alpha_1-\alpha_2.
\]

\subsection{In-Sample Error} \label{sec: uncertainty quantification, in-sample}

\citet*{Cattaneo-Feng-Titiunik_2021_JASA} provide a principled simulation-based method for quantifying the in-sample uncertainty coming from $\bp_T'(\widehat\bbeta-\bbeta_0)$. Let $\bZ=(\bB, \bC)$ and $\bD$ be a non-negative diagonal (scaling) matrix of size $d$, possibly depending on the pre-treatment sample size $T_0$. Since
$\widehat{\bbeta}$ solves \eqref{eq: estimated weight},   $\widehat{\bdelta}:=\bD(\widehat{\bbeta}-\bbeta_0)$ is the optimizer of the centered criterion function:
\[
\widehat{\bdelta}=\underset{\bdelta\in\Delta}{\argmin}\; \big\{\bdelta'\widehat{\bQ}\bdelta-2\widehat{\bgamma}'\bdelta\big\},
\]
where $\widehat{\bQ}=\bD^{-1}\bZ'\bV\bZ\bD^{-1}$,  
$\widehat{\bgamma}'=\bu'\bV\bZ\bD^{-1}$, and $\Delta=\{\bm{h}\in\mathbb{R}^d: \bm{h}=\bD(\bbeta-\bbeta_0),\, \bbeta\in\mathcal{W}\times \mathcal{R}\}$. Recall that the information set conditional on which our prediction intervals are constructed contains $\bB$ and $\bC$. Thus, $\widehat\bQ$ can be taken as fixed, and we need to characterize the uncertainty of $\widehat\bgamma$. 

We construct a simulation-based criterion function accordingly:
\begin{align}
\ell^{\ttb}(\bdelta) = \bdelta'\widehat{\bQ}\bdelta-2(\bG^{\ttb})'\bdelta,\qquad \bG^{\ttb}\thicksim\mathsf{N}(\mathbf{0},\widehat{\bSigma}),
\label{eq: sim unc}
\end{align}
where $\widehat{\bSigma}$ is some estimate of $\bSigma=\V[\widehat\bgamma|\mathscr{H}]$ and $\mathsf{N}(0, \widehat\bSigma)$ represents the normal distribution with mean $\mathbf{0}$ and variance-covariance matrix $\widehat\bSigma$. In practice, the criterion function $\ell^{\ttb}(\cdot)$ can be simulated by simply drawing normal random vectors $\bG^{\ttb}$.

Since the original constraint set $\Delta$ is infeasible, we need to construct a constraint set $\Delta^{\ttb}$  used in simulation that is close to $\Delta$. Specifically, define the distance between a point $\ba\in\mathbb{R}^d$ and 
a set $\Lambda\subseteq\mathbb{R}^d$ by
\[
\mathrm{dist}(\ba, \Lambda)=\inf_{\lambda\in\Lambda}\|\ba-\blambda\|,
\]
where $\|\cdot\|$ is a generic $\ell_p$ vector norm  on $\mathbb{R}^d$ with $p\geq 1$ (e.g., Euclidean norm or $\ell_1$ norm). 
We require
\begin{equation} \label{eq: locally equal}
\mathrm{dist}(\ba, \Delta^{\ttb})\ll\|\ba\|, \quad \forall\,\ba\in\Delta\cap\mathcal{B}(\mathbf{0}, \varepsilon),
\end{equation}
where $\mathcal{B}(\mathbf{0}, \varepsilon)$ is an $\varepsilon$-neighborhood around zero for some $\varepsilon>0$.
In words, every point in the infeasible constraint set $\Delta$ has to be sufficiently \textit{close} to the feasible constraint set used in simulation. We discuss below a principled strategy for constructing $\Delta^\star$, which allows for both linear and non-linear constraints in the feasibility set. Section \ref{sec: uncertainty quantification -- implementation} provides details on how $\Delta^{\ttb}$ is constructed and implemented in the \texttt{scpi} package.

Given the feasible criterion function $\ell^\star(\cdot)$ and constraint set $\Delta^\star$, we let
\begin{align*}
&M_{1,\mathtt{L}} := (\alpha_1/2)\text{-quantile  of }\inf\,\Big\{\bp_T'\bD^{-1}\bdelta: \bdelta\in\Delta^{\ttb},\,\ell^{\ttb}(\bdelta)\leq 0 \Big\}, \qquad\text{and}\\
&M_{1,\mathtt{U}}:=(1-\alpha_1/2)\text{-quantile  of } \sup\,\Big\{\bp_T'\bD^{-1}\bdelta: \bdelta\in\Delta^{\ttb},\,\ell^{\ttb}(\bdelta)\leq 0 \Big\},
\end{align*}
\textit{conditional} on the data. Under mild regularity conditions, for a large class of synthetic control predictands \eqref{eq: estimated weight}, with high probability over $\mathscr{H}$,
\[
\P\big[M_{1,\mathtt{L}}\leq\bp_T'(\widehat{\bbeta}-\bbeta_0)\leq M_{1,\mathtt{U}} \; \big| \; \mathscr{H} \big]\geq 1-\alpha_1,
\]
up to some small loss of the (conditional) coverage probability. Importantly, this conclusion holds whether the data are stationary or non-stationary and whether the model is correctly specified (i.e., $\E[\bu|\mathscr{H}]=0$) or not. If constraints imposed are non-linear, an additional adjustment to this bound may be needed to ensure the desired coverage.

\subsection{Out-of-Sample Error} \label{sec: uncertainty quantification, out-of-sample}

The unobserved random variable $e_T$ in \eqref{eq: SC Y1T(0)} is a single error term in period $T$, which we interpret as the error from out-of-sample prediction, conditional on $\mathscr{H}$. Naturally, in order to have a proper bound on $e_T$, it is necessary to determine certain features of its conditional distribution $F_{e_T}(\mathsf{e}) = \P[e_T \leq \mathsf{e} | \mathscr{H}]$. In this section, we outline principled but agnostic approaches to quantify the uncertainty introduced by the post-treatment unobserved shock $e_T$. Since formalizing the validity of our methods usually requires strong assumptions, we also recommend a generic sensitivity analysis to incorporate out-of-sample uncertainty into the prediction intervals. See Section \ref{sec:sens analysis} and Section \ref{sec: illustration}, in particular Figure \ref{fig:sens analysis} with the corresponding snippet of \texttt{R} code, for further clarifications on how to carry out sensitivity analysis on $e_T$.	

\begin{itemize}[leftmargin=*]
    \item \textbf{Approach 1: Non-Asymptotic bounds}. 
    The starting point is a non-asymptotic probability bound on $e_T$ via concentration inequalities. For example, suppose that $e_T$ is sub-Gaussian conditional on $\mathscr{H}$, i.e., there exists some $\sigma_{\mathscr{H}}>0$ such that
	$\E[\exp(\lambda(e_T-\E[e_T|\mathscr{H}]))|\mathscr{H}]\leq \exp(\sigma_{\mathscr{H}}^2\lambda^2/2)$ a.s. for all $\lambda\in\mathbb{R}$.
	Then, we can take
	\[
	M_{2,\mathtt{L}}:=\E[e_T|\mathscr{H}]-\sqrt{2\sigma_{\mathscr{H}}^2\log(2/\alpha_2)} 
	\qquad\text{and}\qquad
	M_{2,\mathtt{U}}:=\E[e_T|\mathscr{H}]+\sqrt{2\sigma_{\mathscr{H}}^2\log(2/\alpha_2)}.
	\]
	In practice, the conditional mean $\E[e_T|\mathscr{H}]$ and the sub-Gaussian parameter $\sigma_{\mathscr{H}}$ can be parameterized and/or estimated using the pre-treatment residuals.

	\item \textbf{Approach 2: Location-scale model}. Suppose that $e_T=\E[e_T|\mathscr{H}]+(\V[e_T|\mathscr{H}])^{1/2}\varepsilon_T$ with $\varepsilon_T$ statistically independent of $\mathscr{H}$. This setting imposes restrictions on the distribution of $e_T|\mathscr{H}$, but allows for a much simpler tabulation strategy. Specifically, we can set the lower bound and upper bound on $e_T$ as follows:
	\[
	M_{2, \mathtt{L}}=\E[e_T|\mathscr{H}]+(\V[e_T|\mathscr{H}])^{1/2}\mathfrak{c}_\varepsilon(\alpha_2/2) 
	\quad \text{and} \quad 
	M_{2, \mathtt{U}}=\E[e_T|\mathscr{H}]+(\V[e_T|\mathscr{H}])^{1/2}\mathfrak{c}_\varepsilon(1-\alpha_2/2),
	\]	 
	where $\mathfrak{c}_\varepsilon(\alpha_2/2)$ and $\mathfrak{c}_\varepsilon(1-\alpha_2/2)$ are $\alpha_2/2$  and $(1-\alpha_2/2)$ quantiles of $\varepsilon_T$, respectively. In practice, $\E[e_T|\mathscr{H}]$ and $\V[e_T|\mathscr{H}]$ can be parametrized and estimated using the pre-intervention residuals, or perhaps tabulated using auxiliary information. Once such estimates are available, the appropriate quantiles can be easily obtained using the standardized (estimated) residuals.
	 
	\item \textbf{Approach 3: Quantile regression}. Another strategy to bound $e_T$ is to determine the $\alpha_2/2$ and $(1-\alpha_2/2)$ conditional quantiles of $e_T|\mathscr{H}$, that is, 
	\[
    M_{2,\mathtt{L}} := (\alpha_2/2)\text{-quantile  of } e_T | \mathscr{H} 
    \qquad\text{and}\qquad
    M_{2,\mathtt{U}}:=(1-\alpha_2/2)\text{-quantile  of } e_T|\mathscr{H}. 
    \]
	Consequently, we can employ quantile regression methods to estimate those quantities using pre-treatment data.
\end{itemize}

Using any of the above methods, we have the following probability bound on $e_T$:
\[
\P\big[M_{2,\mathtt{L}}\leq e_T\leq M_{2,\mathtt{U}} \; \big| \; \mathscr{H} \big]\geq 1-\alpha_2.
\]

\subsection{Implementation}\label{sec: uncertainty quantification -- implementation}
We now discuss the implementation details. The function \texttt{scpi()}, through various options, allows the user to specify different approaches to quantify in-sample and out-of-sample uncertainty based on the methods described above. Most importantly, \texttt{scpi()} permits modelling separately the in-sample error $\bp_T'(\widehat\bbeta-\bbeta_0)$ and the out-of-sample error $e_T$. In addition, the user can provide bounds on them manually with the options \texttt{w.bounds} and \texttt{e.bounds}, respectively, which can be useful for sensitivity analysis in empirical applications.

\subsubsection*{Modelling In-Sample Uncertainty}\label{sec: how to in sample}

In-sample uncertainty stems from the prediction of $\bp_T'(\widehat\bbeta-\bbeta_0)$, and its quantification reduces to determining $M_{1,\mathtt{L}}$ and $M_{1,\mathtt{U}}$. We first review the methodological proposals for constructing the constraint set $\Delta^\star$ used in simulation discussed in \citet*{Cattaneo-Feng-Palomba-Titiunik_2022_wp}, and then present the main procedure for constructing bounds on the in-sample error.

\bigskip
\noindent\uline{Constructing $\Delta^\star$}. Our in-sample uncertainty quantification requires the centered and scaled constraint feasibility set $\Delta$ to be locally identical to (or, at least, well approximated by) the constraint set $\Delta^\star$ used in simulation described in \eqref{eq: sim unc}, in the sense of \eqref{eq: locally equal}. Suppose that 
$$
\mathcal{W}\times\mathcal{R}=\Big\{\bbeta\in\mathbb{R}^d:\bm{m}_{\mathtt{eq}}(\bbeta)=\bm{0},\bm{m}_{\mathtt{in}}(\bbeta)\leq \bm{0}\Big\},
$$
where $\bm{m}_{\teq}(\cdot)\in\mathbb{R}^{d_{\teq}}$ and $\bm{m}_{\tin}(\cdot)\in\mathbb{R}^{d_{\tin}}$ and denote the $j$th constraint in $\bm{m}_{\tin}(\cdot)$ as $m_{\tin,j}(\cdot)$. Given tuning parameters $\varrho_j>0$, $j=1, \cdots, d_{\tin}$, let $\mathcal{B}$ be the set of indices for the inequality constraints such that $m_{\tin, j}(\hat{\bbeta})>-\varrho_j$. Then, we construct $\Delta^\star$ as
\[
\Delta^\star=\Big\{
\bD(\bbeta-\widehat{\bbeta}):\bm{m}_{\teq}(\bbeta)=\bm{0}, m_{\tin,j}(\bbeta)\leq 
m_{\tin,j}(\widehat{\bbeta}) \text{ for } j\in\mathcal{B}, \text{ and }
m_{\tin,l}(\bbeta)\leq \bm{0} \text{ for } l\notin\mathcal{B}
\Big\}.
\]
In practice, we need to choose possibly heterogeneous parameters $\varrho_j$, $j=1,\ldots, d_\tin$, for different inequality constraints. Our proposed choice of $\varrho_j$ is
\[
\varrho_j:=\Big\|\frac{\partial}{\partial\bbeta}m_{\tin,j}(\widehat{\bbeta})\Big\|_1\times
\varrho, \quad j=1,\ldots,d_\tin,
\]
for some parameter $\varrho$ 
where $\|\cdot\|_1$ denotes the $\ell_1$-norm.  
We estimate $\varrho$ according to the following formula if $M=1$: 
\[ \varrho = \mathcal{C} \frac{\log(T_0)^c}{T_0^{1/2}},\]
where $c=1/2$ if the data are i.i.d. or weakly dependent, and $c=1$ if $\bA$ and $\bB$ form a cointegrated system, while $\mathcal{C}$ is one of the following:
\[\mathcal{C}_1 = \frac{\widehat{\sigma}_{u}}{\min_{1\leq j \leq J}\widehat{\sigma}_{b_j}}, \qquad
  \mathcal{C}_2 = \frac{\max_{1\leq j \leq J}\widehat{\sigma}_{b_j}\widehat{\sigma}_{u}}{\min_{1\leq j \leq J}\widehat{\sigma}^2_{b_j}}, \qquad
  \mathcal{C}_3 = \frac{\max_{1\leq j \leq J}\widehat{\sigma}_{b_ju}}{\min_{1\leq j \leq J}\widehat{\sigma}^2_{b_j}},
 \]
with $\mathcal{C}_1$ as the default. $\widehat\sigma_{b_j,u}$ is the estimated (unconditional) covariance between the pseudo-true residual $\bu$ and the feature of the $j$th control unit $\bB_{j,1}$, and $\widehat\sigma_u$ and $\widehat\sigma_{b_j}$ are the estimated (unconditional) standard deviation of, respectively, $\bu$ and $\bB_{j,1}$. In the case of multiple features ($M>1$), the package employs the same construction above after stacking the data.

\bigskip
\noindent\uline{Degrees-of-Freedom Correction}. Our uncertainty quantification strategy requires an estimator of the conditional variance $\V[\bu | \mathscr{H}]$, which may rely on the effective degrees of freedom $\mathsf{df}$ of the synthetic control method. In general, there exists no exact correspondence between the degrees of freedom and the number of parameters in a fitting model \citep{Ye_1998_JASA}. Therefore, the estimated degrees of freedom $\widehat{\mathsf{df}}$ are defined according to the chosen constraint sets for $\bbeta$ underlying the estimation procedure in \eqref{eq: estimated weight}:

    \begin{itemize}[leftmargin=*]
        \item \textbf{OLS}. $\widehat{\mathsf{df}}=J+KM$.
        
        \item \textbf{Lasso}. Following \citet*{Zou-Hastie-Tibshirani_2007_AS}, 
        an unbiased and consistent estimator of $\mathsf{df}$ is $\widehat{\mathsf{df}} = \sum_{j=1}^J\I(\widehat{w}_j>0)+KM$
        where $\widehat{w}_j$ is the $j$th element of the constructed weights $\widehat{\bw}$.

        \item \textbf{Simplex}. Following the discussion for Lasso, %
        $\widehat{\mathsf{df}} = \sum_{j=1}^J\I(\widehat{w}_j>0)-1+KM$.
        
        \item \textbf{Ridge}. Let $s_1\geq s_2\geq \cdots \geq s_{J}\geq 0$ be singular values of $\bB$ and $\lambda$ be the complexity parameter of the corresponding Lagrangian Ridge problem, which satisfies $\lambda\widehat{\bw}=\bB'(\bA-\bB\widehat{\bw})$. Then, following \citet*{Friedman-Hastie-Tibshirani_2001_Elements}, $\widehat{\mathsf{df}} =\sum_{j=1}^J\frac{s_j^2}{s_j^2+\lambda}+KM$.
    \end{itemize} 

\bigskip
\noindent\uline{Main procedure}. Given the constraint set $\Delta^\star$, the main procedure for computing the upper and lower bounds on the in-sample error is as follows:
\begin{enumerate}[leftmargin=*,label=Step \arabic*.]

  \item \textit{Estimation of conditional moments of $\bu$.} To estimate $\bSigma$ and to simulate the criterion function \eqref{eq: sim unc} we need an estimate of $\V[\widehat\bgamma |\mathscr{H}]$ which, in turn, depends on the conditional moments of $\bu$. To estimate such moments, the user needs to specify three things:
  \begin{enumerate}[label=\roman*)]
    \item  whether the model is misspecified or not, via the option \texttt{u.missp}.
    \item  how to model $\bu$, via the options \texttt{u.order}, \texttt{u.lags}, and \texttt{u.design}.
    \item  an estimator of $\V[\bu|\mathscr{H}]$, via the option \texttt{u.sigma}.
  \end{enumerate}
  Given the constructed weights $\widehat{\bw}=(\widehat{w}_1, \cdots, \widehat{w}_J)'$, define regularized weights $\widehat{\bw}^\star=(\widehat{w}_1^\star, \cdots, \widehat{w}_J^\star)'$ with $\widehat{w}_j^\star=\widehat{w}_j\mathds{1}(\widehat{w}_j>\varrho)$ for the tuning parameter $\varrho$ specified  previously. 
  Let $\bB^\star=\diag(\bB^{\star}_1,\bB^{\star}_2,\ldots,\bB^{\star}_M)$, where $\bB^\star_l$ denotes the matrix composed of the columns of $\bB_l$ with non-zero regularized weight $\widehat{w}_j^\star$ only. If the option \texttt{cointegrated.data} in \texttt{scdata()} is set to be \texttt{TRUE}, rather than the columns of $\bB_l$, we take the first difference of the columns of $\bB_l$. If the user inputs \texttt{u.missp = FALSE}, then it is assumed that $\mathbb{E}[\bu|\mathscr{H}] = 0$, whereas if \texttt{u.missp = TRUE} (default), then $\mathbb{E}[\bu|\mathscr{H}]$ needs to be estimated.
  
  The unknown conditional expectation $\mathbb{E}[\bu|\mathscr{H}]$ is estimated using the fitted values of a flexible linear-in-parameters regression of $\widehat\bu=\bA-\bB\widehat\bw-\bC\widehat\br$ on a design matrix $\bD_{\bu}$, which can be provided directly with the option \texttt{u.design} or by specifying the lags of $\bB^\star$ (\texttt{u.lags}) and/or the order of the fully interacted polynomial in $\bB^\star$ (\texttt{u.order}).
  
  For example, if the user specifies \texttt{u.lags = 1} and \texttt{u.order = 1}, then the design matrix is $\bD_{\bu} = [\bB^\star\;\; \bB^\star_{-1}\;\;\bC]$, where $\bB^\star_{-1}$ indicates the first lag of $\bB^\star$. If, instead, \texttt{u.order = 0} and \texttt{u.lags = 0} are specified, then $\widehat{\mathbb{E}}[\bu|\mathscr{H}] =\overline{\mathbf{u}}\otimes \boldsymbol{\iota}_{T_0}$, where $\overline{\bu}=(\overline{u}_{1},\overline{u}_{2},\ldots,\overline{u}_{M})'$ with $\overline{u}_l= T_0^{-1}\sum_{t=1}^{T_0}\widehat u_{t,l}$, $\boldsymbol{\iota}_{\nu}$ is a $\nu\times 1 $ vector of ones, and 
  $\otimes$ denotes the Kronecker product. 
  
  The conditional variance of $\bu$ is estimated as 
  \[\widehat{\V}[\bu|\mathscr{H}]= \diag\Big(
  \mathtt{vc}_1(\widehat{u}_{1,1}-\widehat{\mathbb{E}}[u_{1,1}|\mathscr{H}])^2,
  \cdots,
  \mathtt{vc}_{T_0\cdot M}(\widehat{u}_{T_0,M}-\widehat{\mathbb{E}}[u_{T_0,M}|\mathscr{H}])^2\Big)\]
  where  $\mathtt{vc}_i$, $i=1, \cdots, T_0\cdot M$ is a sequence of  variance-correction constants, which can be chosen among the well-known family of  heteroskedasticity-robust variance-covariance estimators through the option \texttt{u.sigma}. In particular, the package currently allows for five choices:
  \[\mathtt{vc}_i^{(0)} = 1,\quad \mathtt{vc}_i^{(1)} = \frac{T_0\cdot M}{T_0\cdot M-\mathsf{df}},\quad \mathtt{vc}_i^{(2)} = \frac{1}{1-\bL_{ii}},\quad \mathtt{vc}_i^{(3)} =\frac{1}{\left(1-\bL_{ii}\right)^2} ,\quad \mathtt{vc}_i^{(4)} = \frac{1}{\left(1-\bL_{ii}\right)^{\delta_i}}\]
  with $\bL_{ii}$ being the $i$-th diagonal entry of the leverage matrix $\bL := \bZ(\bZ'\bV\bZ)^{-1}\bZ'\bV$, $\delta_i =\min\{4,T_0\cdot M\cdot \bP_{ii} / \mathsf{df}\}$, and $\mathsf{df}$ is a degrees-of-freedom correction factor, whose estimation has been explained before. 

  \item \textit{Estimation of $\bSigma$}. The estimator of $\bSigma$ is $\widehat{\bSigma} = (\bZ'\bV) \widehat{\V}[\bu|\mathscr{H}](\bV \bZ)$.
  
  \item \textit{Simulation}. The criterion function $\ell^{\ttb}(\bdelta)$ in \eqref{eq: sim unc} is simulated by drawing i.i.d. random vectors from the Gaussian distribution $\mathsf{N}(0,\widehat{\bSigma})$, conditional on the data.
  
  \item \textit{Optimization}. Let $\ell_{(s)}^\star(\bdelta)$ denote the criterion function corresponding to the $s$-th draw from $\mathsf{N}(0,\widehat{\bSigma})$. For each draw $s$, we solve the following constrained problems:
  \begin{align}\label{eq:in sam obj}
    l_{(s)}:=\inf_{\bdelta \in \Delta^\star,\, \ell_{(s)}^{\ttb}(\bdelta)\leq 0} \bp_T'\bD^{-1}\bdelta \qquad\text{and}\qquad
    u_{(s)}:=\sup_{\bdelta \in \Delta^\star,\, \ell_{(s)}^{\ttb}(\bdelta)\leq 0} \bp_T'\bD^{-1}\bdelta,
  \end{align}
  where $\Delta^\star$ is constructed as explained previously.
  
  \item \textit{Estimation of $M_{1,\mathtt{L}}$ and $M_{1,\mathtt{U}}$}. Step 4 is repeated $S$ times, where $S$ can be specified with the option \texttt{sims}. Then, $M_{1,\mathtt{L}}$ is the $(\alpha_1/2)-$quantile of $\{l_{(s)}\}_{s=1}^S$ and $M_{1,\mathtt{U}}$ is the $(1-\alpha_1/2)-$quantile of $\{u_{(s)}\}_{s=1}^S$. The level of $\alpha_1$ can be chosen with the option $\texttt{u.alpha}$.
\end{enumerate}

\bigskip
\noindent\uline{Execution Speed and Parallelization}. Steps 3 and 4 of the procedure above are the most computationally intensive and we optimize them in two ways. First, to solve the optimization problem in \eqref{eq:in sam obj}, \texttt{scpi} relies on \texttt{ECOS}, an efficient solver for conic problems \citep*{Domahidi-Chu-Boyd_2013_ECOS, Fu-Narasimhan-Boyd_2020_JSS}. See \citet*{Cattaneo-Feng-Palomba-Titiunik_2022_wp} for more details on how to cast the different constrained SC methods into conic optimization problems. To give the reader a sense of the speed improvement, Table \ref{tab:speed comparison} compares the execution speed of the conic solver we rely on (first column) with other two popular optimizers in \texttt{R}. The first row of the table reports the median computation time of each optimizer, whereas the second row shows the inter-quartile range. On the one hand, using a conic solver in place of a solver for more generic optimization programs (like \texttt{nloptr}) makes our software 4 times faster. On the other hand, our software is tailored to rewrite the SC problem as a conic problem. This gives a 300-fold gain in speed when compared to \texttt{CVXR}, which relies on \texttt{ECOS} but is meant for prototyping generic optimization problems in conic form. 

\begin{table}[!ht]
  \centering
  \caption{Speed comparison across optimizers (units: milliseconds).}
    \begin{tabular}{lccc}
    \toprule\toprule
          & \texttt{ECOS}  & \texttt{CVXR}  & \texttt{nloptr} \\
    \midrule
    Median & 1.411 & 308.823 & 5.734 \\
    IQR   & [1.387, 1.431] & [301.183, 315.901] & [5.534, 6.148] \\
    \bottomrule\bottomrule
    \end{tabular}%
  \label{tab:speed comparison}%
      \par
	\begin{center}
		\parbox[1]{\textwidth}{\footnotesize \textit{Notes:} The underlying optimization problem is the minimization problem in $\eqref{eq:in sam obj}$, where $\mathcal{W}$ is a simplex-type constraint and $J,KM,$ and $M$ are chosen to replicate the size of the empirical application in Section \ref{sec: illustration}. We evaluate the performance of the function \texttt{scpi} through the \texttt{R} package \texttt{microbenchmark}. This simulation was run using an Apple M2 chip, RAM 8.00 GB.}
	\end{center}    
\end{table}%

Second, \texttt{scpi} can be sped up further by efficient parallelization of the tasks performed through the package \texttt{parallel} which assigns different simulations to different cores. If $\mathtt{N}_{\text{cores}}$ cores are used, the final execution time would be approximately $\mathtt{T}_{\text{exec}}/\mathtt{N}_{\text{cores}}$, where $\mathtt{T}_{\text{exec}}$ is the execution time when a single core is used.

\subsubsection*{Modelling Out-of-Sample Uncertainty}\label{sec:how to out of sample}

To quantify the uncertainty coming from $e_T$, we need to impose some probabilistic structure that allows us to model the distribution $\mathbb{P}[e_T\leq \mathtt{e} |\mathscr{H}]$ and, ultimately, estimate $M_{2,\mathtt{L}}$ and $M_{2,\mathtt{U}}$. We discussed three different alternative approaches: (i) non-asymptotic bounds; (ii) location-scale model; and (iii) quantile regression. The user can choose the preferred way of modeling $e_T|\mathscr{H}$ by setting the option \texttt{e.method} to either \texttt{`gaussian'}, \texttt{`ls'}, or \texttt{`qreg'}.

The user can also choose the information used to estimate (conditional) moments or quantiles of $e_T|\mathscr{H}$. Practically, we allow the user to specify a design matrix $\bD_{\be}$ that is then used to run the appropriate regressions depending on the approach requested. By default, we set $\bD_{\be} = [\bB^\star_1\;\; \bC_1]$. Alternatively, the matrix $\bD_{\be}$ can be provided directly through the option \texttt{e.design} or by specifying the lags of $\bB^\star_1$ (\texttt{e.lags}) and/or the order of the fully interacted polynomial in $\bB^\star_1$ (\texttt{e.order}). If the user specifies \texttt{e.lags = 0} and \texttt{e.order = 2}, then $\bD_{\be}$ contains $\bB^\star_1$, $\bC_1$, and all the unique second-order terms generated by the interaction of the columns of $\bB^\star_1$. If instead \texttt{e.order = 0} and \texttt{e.lags = 0} are set, then $\widehat{\mathbb{E}}[e_T|\mathscr{H}]$ and $\widehat{\mathbb{V}}[e_T|\mathscr{H}]$ are estimated using the sample average and the sample variance of $e_T$ using the pre-intervention data. Recall that if the option \texttt{cointegrated.data} is set to \texttt{TRUE}, $\bB^\star_1$ is formed using the first differences of the columns in $\bB_1$. Finally, the user can specify $\alpha_2$ with the option \texttt{e.alpha}.

\subsection{Simultaneous Prediction Intervals} \label{sec:simultaneous pi}
Up to this point, we focused on prediction intervals that possess high coverage for the individual treatment effect in \textit{each} period. However, it may be desirable to have prediction intervals that have high \textit{simultaneous} coverage for several periods, usually known as \textit{simultaneous prediction intervals} in the literature. In other words, our final goal is to construct a sequence of intervals $\{\mathcal{I}_t: T_0+1\leq t\leq T_0+L\}$ for some $1\leq L \leq T_1$ such that with high probability over $\mathscr{H}$,
\[
  \P\big[\tau_{t}\in\mathcal{I}_t, \text{ for all } T_0+1\leq t\leq T_0+L\,\big|\, \mathscr{H}\big]\geq 1-\alpha_1-\alpha_2.
\]

To construct such intervals, we need to generalize the procedures described above to quantify the in-sample error (Section \ref{sec: uncertainty quantification, in-sample}) and the out-of-sample error (Section \ref{sec: uncertainty quantification, out-of-sample}).

With regard to the in-sample uncertainty, we handle two separate cases. On the one hand, if the constraints in $\Delta$ are linear (e.g., simplex or lasso), then 
\begin{align*}
&M_{1,\mathtt{L}} := (\alpha_1/2)\text{-quantile  of }\inf\,
\Big\{\bp_t'\bD^{-1}\bdelta:\; \bdelta\in\Delta^{\ttb},\;\ell^{\ttb}(\bdelta)\leq 0,\; T_0+1\leq t\leq T_0+L\Big\} \;\;\text{and}\\
&M_{1,\mathtt{U}}:=(1-\alpha_1/2)\text{-quantile  of } \sup\,\Big\{\bp_t'\bD^{-1}\bdelta:\; \bdelta\in\Delta^{\ttb},\;\ell^{\ttb}(\bdelta)\leq 0, \;T_0+1\leq t\leq T_0+L\Big\},
\end{align*}
which guarantees that with high probability over $\mathscr{H}$
\[
  \P\big[M_{1,\mathtt{L}}\leq \bp_{t}'(\bbeta_0-\widehat{\bbeta})\leq M_{1,\mathtt{U}}, \text{ for all } T_0+1\leq t\leq T_0+L\,\big|\, \mathscr{H}\big]\geq 1-\alpha_1.
\]
On the other hand, if $\Delta$ includes non-linear constraints (e.g., constraints involving the $\ell_2$ norm), it is necessary to decrease the lower bound $M_{1,\mathtt{L}}$ and increase the upper bound $M_{1,\mathtt{U}}$ by some quantity $\varepsilon_{\Delta,t}>0$ for each $T_0+1\leq t\leq T_0+L$. To give an example of what $\varepsilon_{\Delta,t}$ looks like, in the case of ridge-type constraints we have
\[
\varepsilon_{\Delta,t}=\|\bp_t\|_1\times (2\|\widehat{\bbeta}\|_2)^{-1}\times \varrho^2,
\]
and see \citet*{Cattaneo-Feng-Palomba-Titiunik_2022_wp} for more general cases. With regard to the out-of-sample uncertainty, our proposed strategy is a generalization of ``Approach 1'' in Section \ref{sec: uncertainty quantification, out-of-sample}: find $M_{2,\mathtt{L},t}$ and $M_{2,\mathtt{U},t}$ such that with high probability over $\mathscr{H}$,
\[
  \P\big[M_{2,\mathtt{L},t}\leq e_t\leq M_{2,\mathtt{U},t}, \text{ for all } T_0+1\leq t\leq T_0+L\,\big|\, \mathscr{H}\big]\geq 1-\alpha_2.
\] 
Suppose that each $e_{t}$, $T_0+1\leq t\leq T_0+L$, is sub-Gaussian conditional on $\mathscr{H}$ (not necessarily independent over $t$) with sub-Gaussian parameters $\sigma_{\mathscr{H},t}\leq \sigma_{\mathscr{H}}$ for some $\sigma_{\mathscr{H}}$. Then, we can take
\[
M_{2,\mathtt{L},t}:=\E[e_t|\mathscr{H}]-\sqrt{2\sigma_{\mathscr{H}}^2\log (2L/\alpha_2)}
\qquad\text{and}\qquad
M_{2,\mathtt{U},t}:=\E[e_t|\mathscr{H}]+\sqrt{2\sigma_{\mathscr{H}}^2\log (2L/\alpha_2)}.
\]
We can see that, compared to what we had for ``Approach 1'', there is an extra term, $\sqrt{\log L}$, which makes the simultaneous prediction intervals longer. 

\subsection{Sensitivity Analysis}\label{sec:sens analysis}

While the three approaches for out-of-sample uncertainty quantification described in Section \ref{sec: uncertainty quantification, out-of-sample} are simple and intuitive, their validity requires potentially strong assumptions on the underlying data generating process that links the pre-treatment and post-treatment data. Such assumptions are difficult to avoid because the ultimate goal is to learn about the statistical uncertainty introduced by a single unobserved random variable after the treatment/intervention is deployed, that is, $e_T|\mathscr{H}$ for some $T>T_0$. Without additional data availability, or specific modelling assumptions allowing for transferring information from the pre-treatment period to the post-treatment period, it is difficult to formally construct $M_{2,\mathtt{L}}$ and $M_{2,\mathtt{U}}$ using data-driven methods.

We suggest approaching the out-of-sample uncertainty quantification as a principled sensitivity analysis, using the approaches above as a starting point. Given the formal and detailed in-sample uncertainty quantification described previously, it is natural to progressively enlarge the final prediction intervals by adding additional out-of-sample uncertainty to ask the question: how large does the additional out-of-sample uncertainty contribution coming from $e_T|\mathscr{H}$ need to be in order to render the treatment effect $\tau_T$ statistically insignificant? Using the approaches above, or similar ones, it is possible to construct natural initial benchmarks. For instance, to implement Approach 1, one can use the pre-treatment outcomes or synthetic control residuals to obtain a ``reasonable'' benchmark estimate of the sub-Gaussian parameter $\sigma_{\mathscr{H}}$ and then progressively enlarge or shrink this parameter to check the robustness of the conclusion. Alternatively, in specific applications, natural levels of uncertainty for the outcomes of interest could be available, and hence used to tabulate the additional out-of-sample uncertainty. We illustrate this approach in Section \ref{sec: illustration}.

\section{Empirical Illustration}\label{sec: illustration}

We showcase the features of the package \texttt{scpi} using real data. For comparability purposes, we employ the canonical dataset in the synthetic control literature on the economic consequences of the 1990 German reunification \citep{Abadie_2021_JEL}, and focus on quantifying the causal impact of the German reunification on GDP per capita in West Germany. Thus, we compare the post-reunification outcome for West Germany with the outcome of a synthetic control unit constructed using 16 OECD countries from 1960 to 1990. Using the notation introduced above, we have $T_0 = 31$ and $J=16$. The only feature we exploit to construct the synthetic control is yearly GDP per capita, and we add a constant term for covariate adjustment. Thus $M=1$ and $K=1$, and $\mathcal{R}=\mathbb{R}$. We explore the effect of the reunification from 1991 to 2003, hence $T_1 = 13$. Finally, we treat the time series for West Germany and those countries in the donor pool as a cointegrating system. Given this information, the command \texttt{scdata()} prepares all the matrices needed in the synthetic control framework described above ($\bA$, $\bB$, $\bC$ and $\bP$), and returns an object that must be used as input in either \texttt{scest()} to predict $\widehat{Y}_{1T}(0),$ $T>T_0$, or \texttt{scpi()} to conduct inference on $\widehat{\tau}_{T},$ $T>T_0$.

We first call \texttt{scdata()} to transform any data frame into an object of class ``\texttt{scpi\_data}''.

{\singlespacing
\begin{lstlisting}
# Load data
> data <- scpi_germany
> 
> ## Set parameters for data preparation
> id.var      <- "country"                     # ID variable
> time.var    <- "year"                        # Time variable
> period.pre  <- (1960:1990)                   # Pre-treatment period
> period.post <- (1991:2003)                   # Post-treatment period
> unit.tr     <- "West Germany"                # Treated unit
> unit.co     <- unique(data$country)[-7]      # Donor pool
> outcome.var <- "gdp"                         # Outcome variable
> constant    <- TRUE                          # Include constant term
> cointegrated.data <- TRUE                    # Cointegrated data
>
# Data preparation
> df  <-   scdata(df = data, id.var = id.var, time.var = time.var, 
+                 outcome.var = outcome.var, period.pre = period.pre, 
+                 period.post = period.post, unit.tr = unit.tr, 
+                 unit.co = unit.co, constant = constant, 
+                 cointegrated.data = cointegrated.data)
\end{lstlisting}}

After having prepared the data, the next step involves choosing the desired constraint set $\mathcal{W}$ to construct the vector of weights $\bw$. We consider the canonical synthetic control method and thus search for optimal weights in $\mathcal{W} = \{\bw \in \mathbb{R}^J_+: ||\bw||_1 = 1\}$. Such constraint set is the default in \texttt{scest()} and, consequently, in \texttt{scpi()},  as the latter internally calls the former to construct $\bw$. The snippet below illustrates how to call \texttt{scest()} and reports the results displayed in the console with the \texttt{summary()} method.

{\singlespacing
\begin{lstlisting}
# Estimate SC with a simplex-type constraint (default)
> res.est <- scest(data = df, w.constr = list(name="simplex"))
> summary(res.est)

Synthetic Control Estimation - Setup

Constraint Type:                           simplex
Constraint Size (Q):                       1
Treated Unit:                              West Germany
Size of the donor pool:                    16
Features:                                  1
Pre-treatment period:                      1960-1990
Pre-treatment periods used in estimation:  31
Covariates used for adjustment:            1

Synthetic Control Estimation - Results

Active donors: 6 

Coefficients:
            Weights
Australia     0.000
Austria       0.441
Belgium       0.000
Denmark       0.000
France        0.000
Greece        0.000
Italy         0.177
Japan         0.013
Netherlands   0.059
New Zealand   0.000
Norway        0.000
Portugal      0.000
Spain         0.000
Switzerland   0.036
UK            0.000
USA           0.274
           Covariates
0.constant      0.158
\end{lstlisting}}

The next step is uncertainty quantification using \texttt{scpi()}. In this case, we quantify the in-sample and out-of-sample uncertainty the same way, using $\bB$ and $\bC$ as the conditioning set in both cases. To do so, it suffices to set the order of the polynomial in $\bB$ to 1 (\texttt{u.order <- 1} and \texttt{e.order <- 1})  and not include lags  (\texttt{u.lags <- 0} and \texttt{e.lags <- 0}). Furthermore, by specifying the option \texttt{u.miss <- TRUE}, we take into account that the conditional mean of $\bu$ might differ from 0. This option, together with \texttt{u.sigma <- "HC1"}, specifies the following estimator of $\V[\bu|\mathscr{H}]$:

\[\widehat\V[\bu|\mathscr{H}] =  \diag\Big(
\mathtt{vc}_1^{(1)}(\widehat{\bu}_1 -\widehat\E[\bu_1 |\mathscr{H}])^2,\cdots,
\mathtt{vc}_{T_0}^{(1)}(\widehat{\bu}_{T_0} -\widehat\E[\bu_{T_0} |\mathscr{H}])^2
\Big).
\]

Finally, by selecting \texttt{e.method <- "gaussian"}, we perform out-of-sample uncertainty quantification treating $e_T$ as sub-gaussian conditional on $\bB$ and $\bC$. As a last step, we visualize the constructed synthetic control and compare it with the observed time series for the treated unit, taking advantage of the function \texttt{scplot()}.

{\singlespacing
\begin{lstlisting}
## Quantify uncertainty
> sims     <- 500          # Number of simulations
> u.order  <- 1            # Degree of polynomial in B and C when modelling u
> u.lags   <- 0            # Lags of B to be used when modelling u
> u.sigma  <- "HC1"        # Estimator for the variance-covariance of u
> u.missp  <- TRUE         # If TRUE then the model is treated as misspecified
> e.order  <- 1            # Degree of polynomial in B and C when modelling e
> e.lags   <- 0            # Lags of B to be used when modelling e
> e.method <- "qreg"       # Estimation method for out-of-sample uncertainty
> lgapp    <- "linear"     # Local geometry approximation
> cores    <- 1            # Number of cores to be used by scpi

> set.seed(8894)
> res.pi  <- scpi(data = df, sims = sims, e.method = e.method, e.order = e.order, 
+                 e.lags = e.lags, u.order = u.order, u.lags = u.lags, lgapp = lgapp,
+                 u.sigma = u.sigma, u.missp = u.missp, cores = cores,
+                 w.constr = list(name = "simplex")) 

# Visualize results
> plot <- scplot(result = res.pi, plot.range = (1960:2003),
+                label.xy = list(x.lab = "Year", x.ticks = NULL, e.out = TRUE, 
+                y.lab = "GDP per capita (thousand US dollars)"),
+                event.label = list(lab = "Reunification", height = 10))

> plot <- plot$plot_out + ggtitle("") 
> ggsave(filename = 'germany_unc_simplex.png', plot = plot)
\end{lstlisting}}

\begin{figure}[H]
    \centering
    \caption{Treated and synthetic unit using a simplex-type $\mathcal{W}$ and 90\% prediction intervals}
    \includegraphics[scale=0.8]{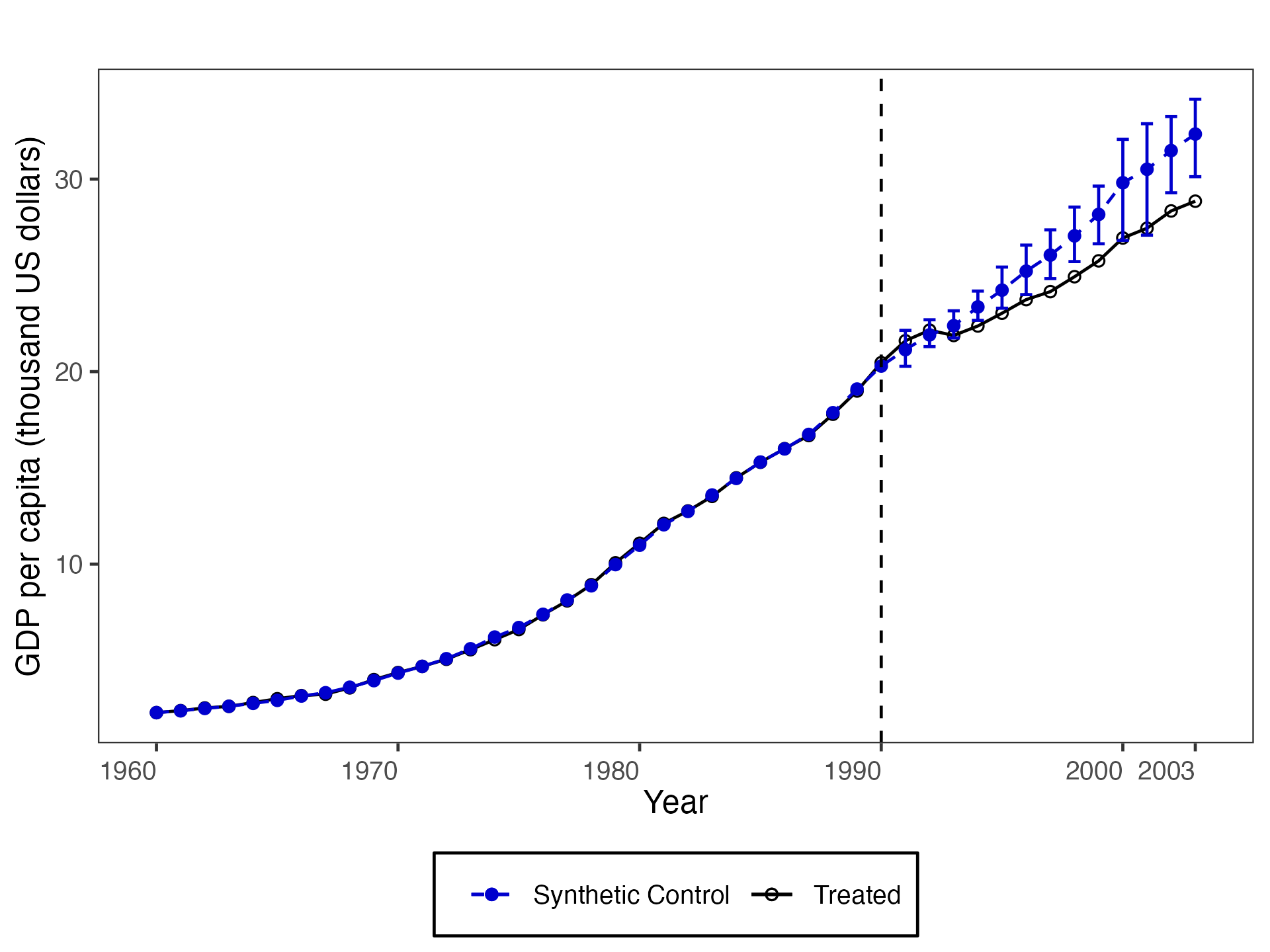}
    \label{fig:germanyunc}
  \par
\begin{center}
	\parbox[1]{\textwidth}{\footnotesize \textit{Notes:} The black line shows the level of the outcome for the treated unit, $Y_{1t}(1)$, $t=1963,\ldots, 2003$, whilst the blue line shows the level of the outcome for the synthetic control, $\widehat{Y}_{1t}(0)$, $t=1963,\ldots,2003$. The blue bars report 90\% prediction intervals for $Y_{1t}(0)$. In-sample uncertainty is quantified by means of 1000 simulations of \eqref{eq:in sam obj}, whereas out-of-sample uncertainty is quantified through sub-Gaussian bounds.}
\end{center}         
\end{figure}

Figure \ref{fig:germanyunc} displays the plot resulting from the \texttt{scplot} call. The vertical bars are 90\% prediction intervals, where the non-coverage error rate is halved between the out-of-sample and the in-sample uncertainty quantification, i.e. $\alpha_1=\alpha_2=0.05$.

\begin{figure}[H]
     \centering
     \caption{Uncertainty quantification with different types of $\mathcal{W}$ using 90\% prediction intervals.}
     \begin{subfigure}[b]{0.48\textwidth}
         \centering
         \includegraphics[width=\textwidth]{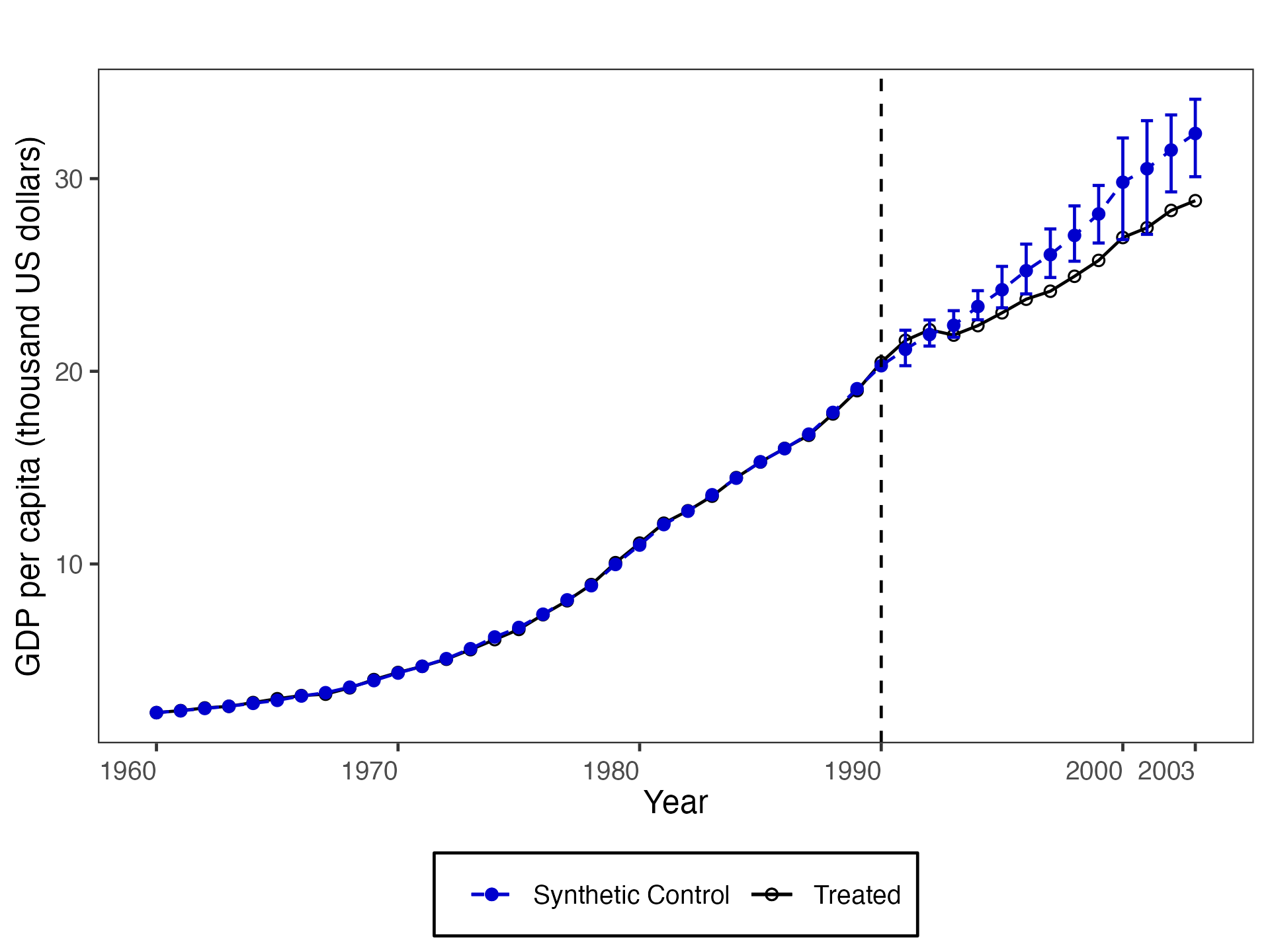}
         \caption{lasso}
         \label{fig:unc lasso}
     \end{subfigure}
     \hfill
     \begin{subfigure}[b]{0.48\textwidth}
         \centering
         \includegraphics[width=\textwidth]{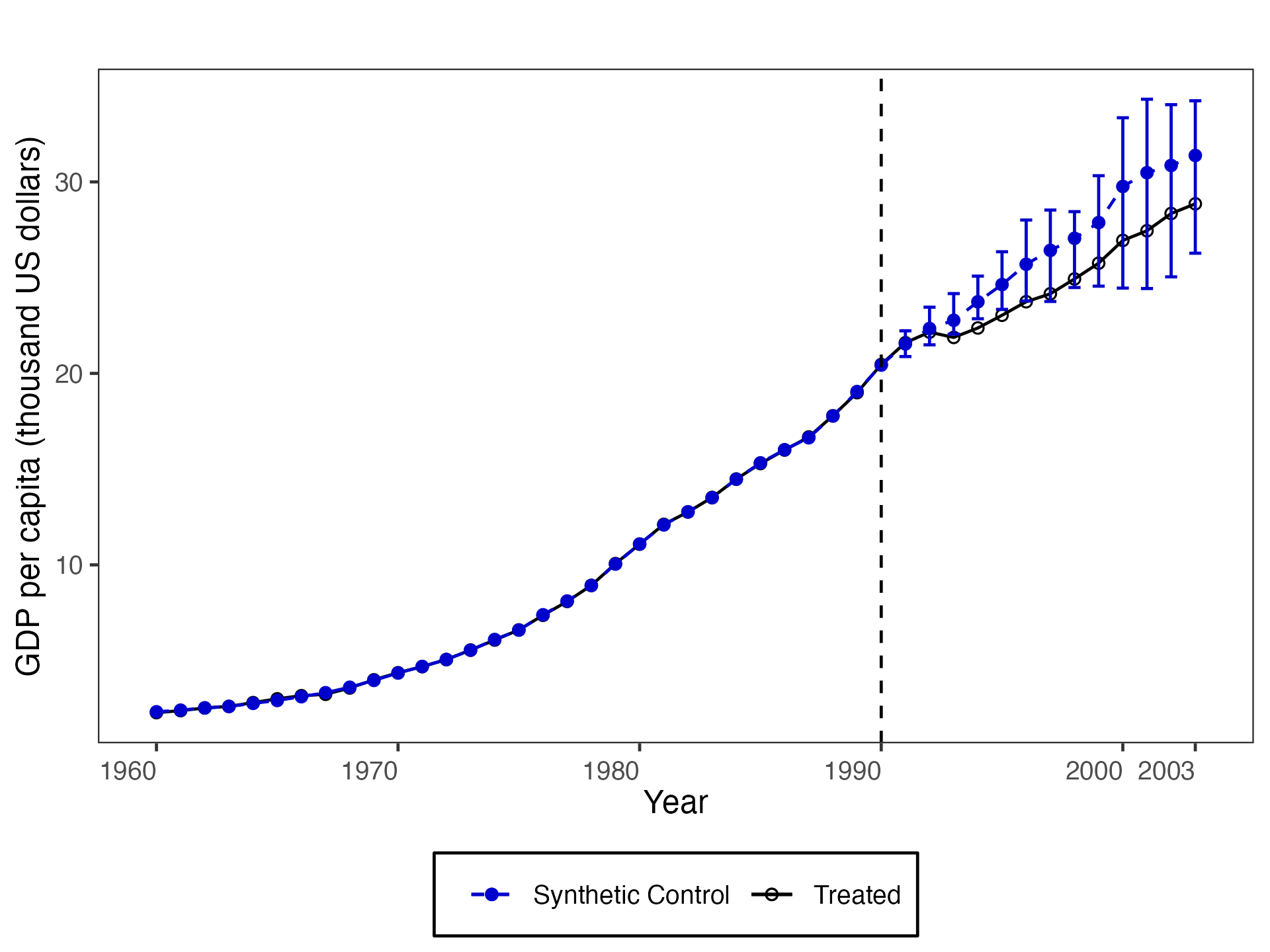}
         \caption{ridge}
         \label{fig:unc ridge}
     \end{subfigure} \\
     \begin{subfigure}[b]{0.48\textwidth}
         \centering
         \includegraphics[width=\textwidth]{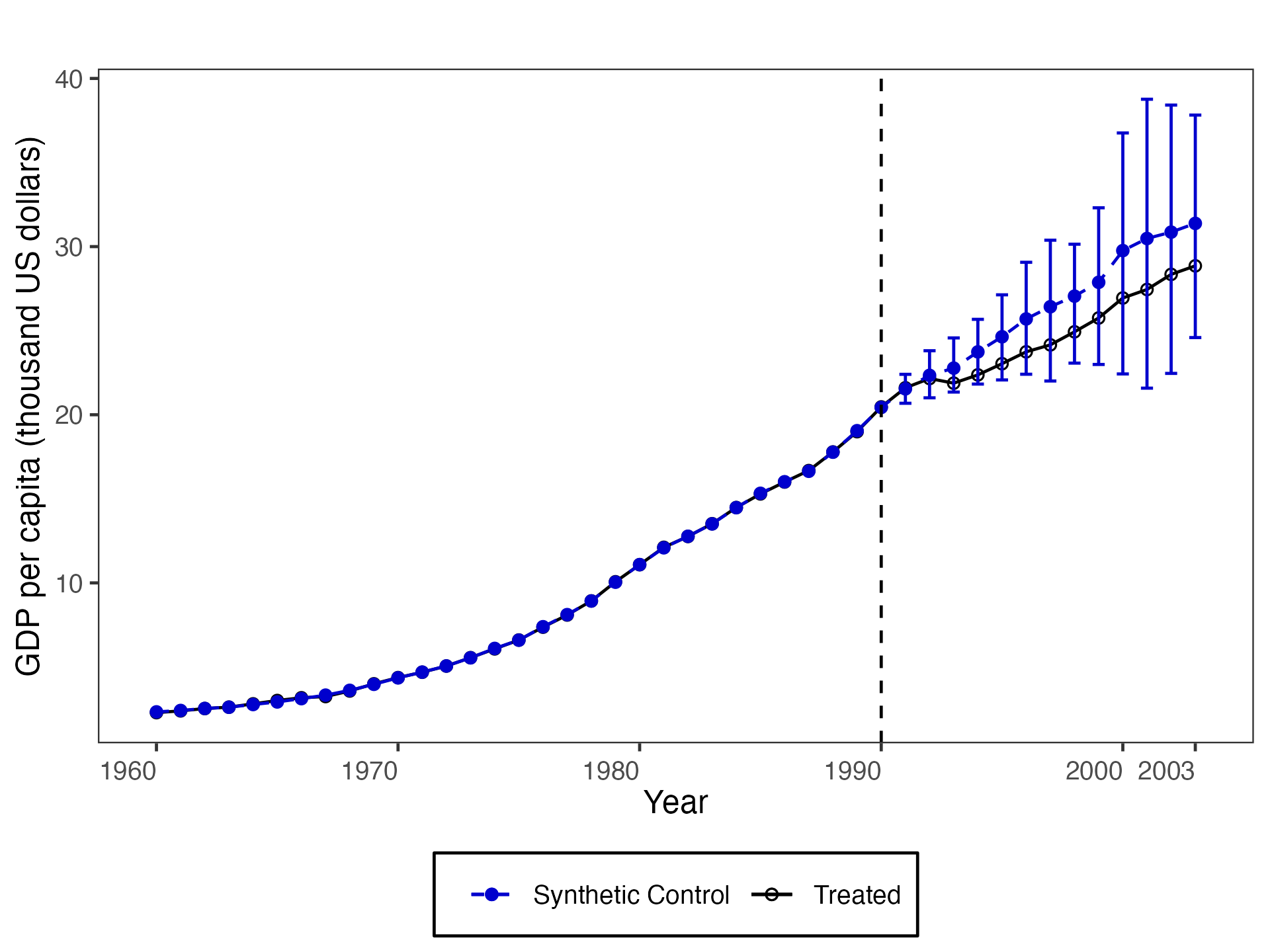}
         \caption{least squares}
         \label{fig:unc ols}
     \end{subfigure}
     \begin{subfigure}[b]{0.48\textwidth}
         \centering
         \includegraphics[width=\textwidth]{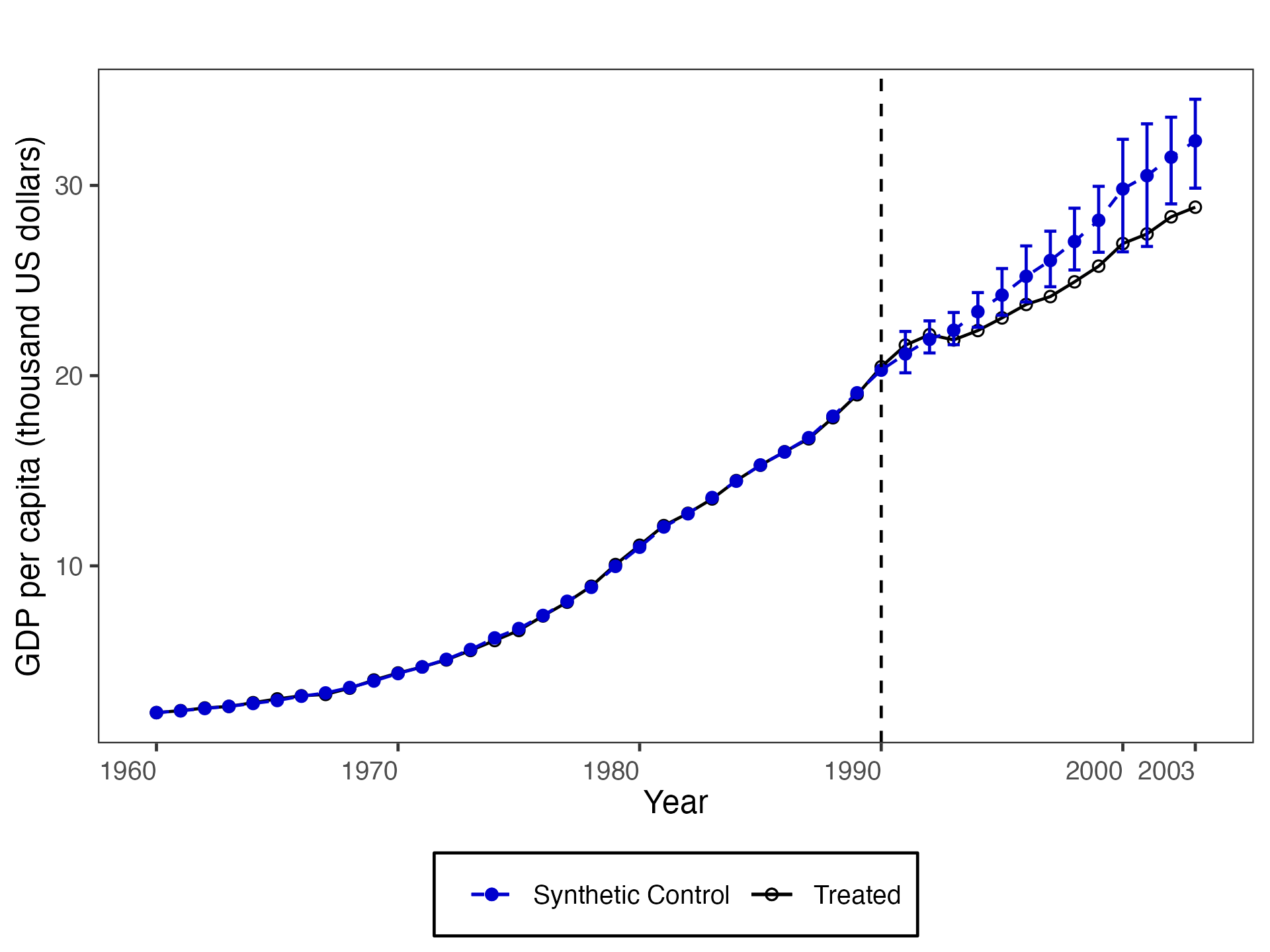}
         \caption{L1-L2}
         \label{fig:unc l1l2}
     \end{subfigure}
  \par
\begin{center}
	\parbox[1]{\textwidth}{\footnotesize \textit{Notes:} The black lines show the level of the outcome for the treated unit, $Y_{1t}(1)$, $t=1963,\ldots, 2003$, whilst the blue lines show the level of the outcome for the synthetic control, $\widehat{Y}_{1t}(0)$, $t=1963,\ldots,2003$. The blue bars report 90\% prediction intervals for $Y_{1t}(0)$. In-sample uncertainty is quantified by means of 1000 simulations of \eqref{eq:in sam obj}, whereas out-of-sample uncertainty is quantified through sub-Gaussian bounds. In panel (b), $Q=0.906$, whereas in panel (d) $Q=1,Q_2=0.906$.}
\end{center}     
 \end{figure}

We also conduct the same exercise using different choices of  $\mathcal{W}$ (see Table \ref{tab:pop_constr}). In particular, we construct weights and compute prediction intervals using four other specifications: (\textit{i}) a lasso-type constraint (Figure \ref{fig:unc lasso}), (\textit{ii}) a ridge-type constraint (Figure \ref{fig:unc ridge}), (\textit{iii}) no constraint (Figure \ref{fig:unc ols}), and (\textit{iv}) an L1-L2 constraint.

{\singlespacing
\begin{lstlisting}
# Comparison of different constraint sets for the weights
> methods <- c("lasso", "ols", "ridge", "L1-L2")

> for (method in methods) {
>   if (method %in%  c("ridge", "L1-L2")) lgapp <- "generalized"
>   set.seed(8894)
>   res.pi  <- scpi(data = df, sims = sims, e.method = e.method, e.order = e.order, 
+                   e.lags = e.lags, u.order = u.order, u.lags = u.lags, lgapp = lgapp,
+                   u.sigma = u.sigma, u.missp = u.missp, cores = cores, 
+                   w.constr = list(name = method)) 

    # Visualize results
>   plot <- scplot(result = res.pi, plot.range = (1960:2003),
+                  label.xy = list(x.lab = "Year", x.ticks = NULL, e.out = TRUE, 
+                  y.lab = "GDP per capita (thousand US dollars)"),
+                  event.label = list(lab = "Reunification", height = 10))

>   plot <- plot$plot_out + ggtitle("") 
>   ggsave(filename = paste0('germany_unc_',method,'.png'), plot = plot)
}
\end{lstlisting}}

\begin{figure}[H]
    \centering
    \caption{Sensitivity analysis on out-of-sample uncertainty with sub-Gaussian bounds.}
    \includegraphics[scale=0.6]{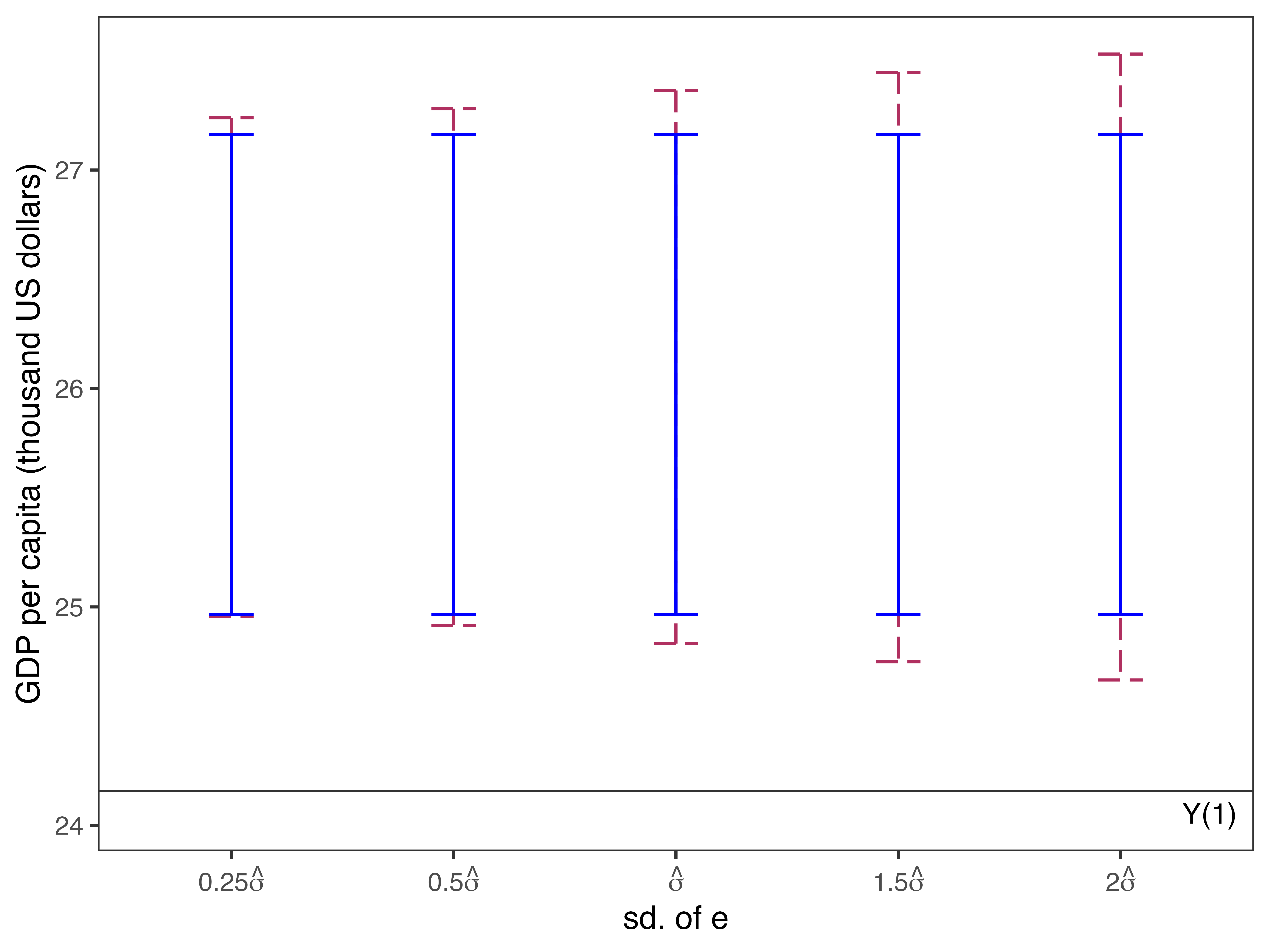}
    \label{fig:sens analysis}
      \par
	\begin{center}
		\parbox[1]{\textwidth}{\footnotesize \textit{Notes:} The black horizontal line shows the level of the outcome for the treated unit in 1997, $Y_{1t}(1)$ for $t=1997$. The blue bars report 95\% prediction intervals for $Y_{1t}(0)$, $t=1997$, that only take into account in-sample uncertainty. The red dashed bars adds the out-of-sample uncertainty to obtain 90\% prediction intervals.}
	\end{center}      
\end{figure}

Section \ref{sec:sens analysis} clarified the need for some additional sensitivity analysis when it comes to out-of-sample uncertainty quantification. Figure \ref{fig:sens analysis} shows the impact of shrinking and enlarging $\widehat\sigma_{\mathscr{H}}$ on the prediction intervals for $Y_{1t}(0)$, $t=1997$, when we assume that $e_t$ is sub-Gaussian conditional on $\mathscr{H}$. As shown in the figure, the predicted treatment effect $\widehat\tau_{\mathtt{1997}}$ remains different from zero with high probability over $\mathscr{H}$ even doubling $\widehat\sigma_\mathscr{H}$ .


Finally, the package offers the possibility to match the treated unit and the synthetic unit using multiple features and the possibility to compute simultaneous prediction intervals. If we want to match West Germany and the synthetic unit not only on GDP per capita but also on trade openness ($M=2$) and include joint prediction intervals, we can simply modify the object \texttt{scpi\_data} as follows.

{\singlespacing
\begin{lstlisting}
## Data preparation
df  <-   scdata(df = data, id.var = id.var, time.var = time.var, 
                outcome.var = outcome.var, period.pre = period.pre, 
                period.post = period.post, unit.tr = unit.tr, 
                features = c("gdp","trade"), cov.adj = list(c("constant")),
                cointegrated.data = cointegrated.data, unit.co = unit.co)
\end{lstlisting}}

\begin{figure}[H]
     \centering
     \caption{Uncertainty quantification with different types of $\mathcal{W}$ using 90\% prediction intervals (2 features).}
     \label{fig:multifeatures}
     \begin{subfigure}[b]{0.48\textwidth}
         \centering
         \includegraphics[width=\textwidth]{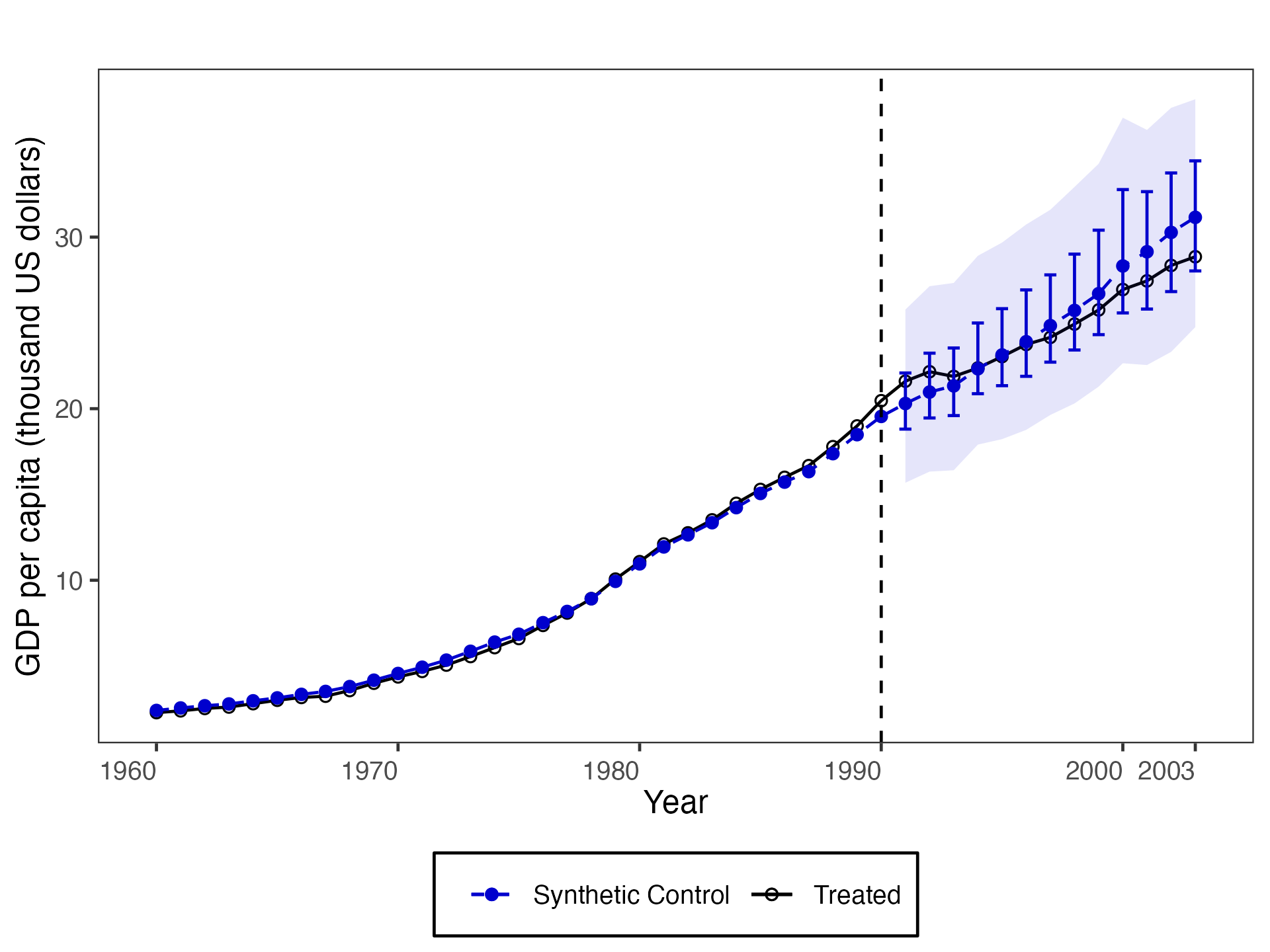}
         \caption{simplex}
     \end{subfigure}\hfill
     \begin{subfigure}[b]{0.48\textwidth}
         \centering
         \includegraphics[width=\textwidth]{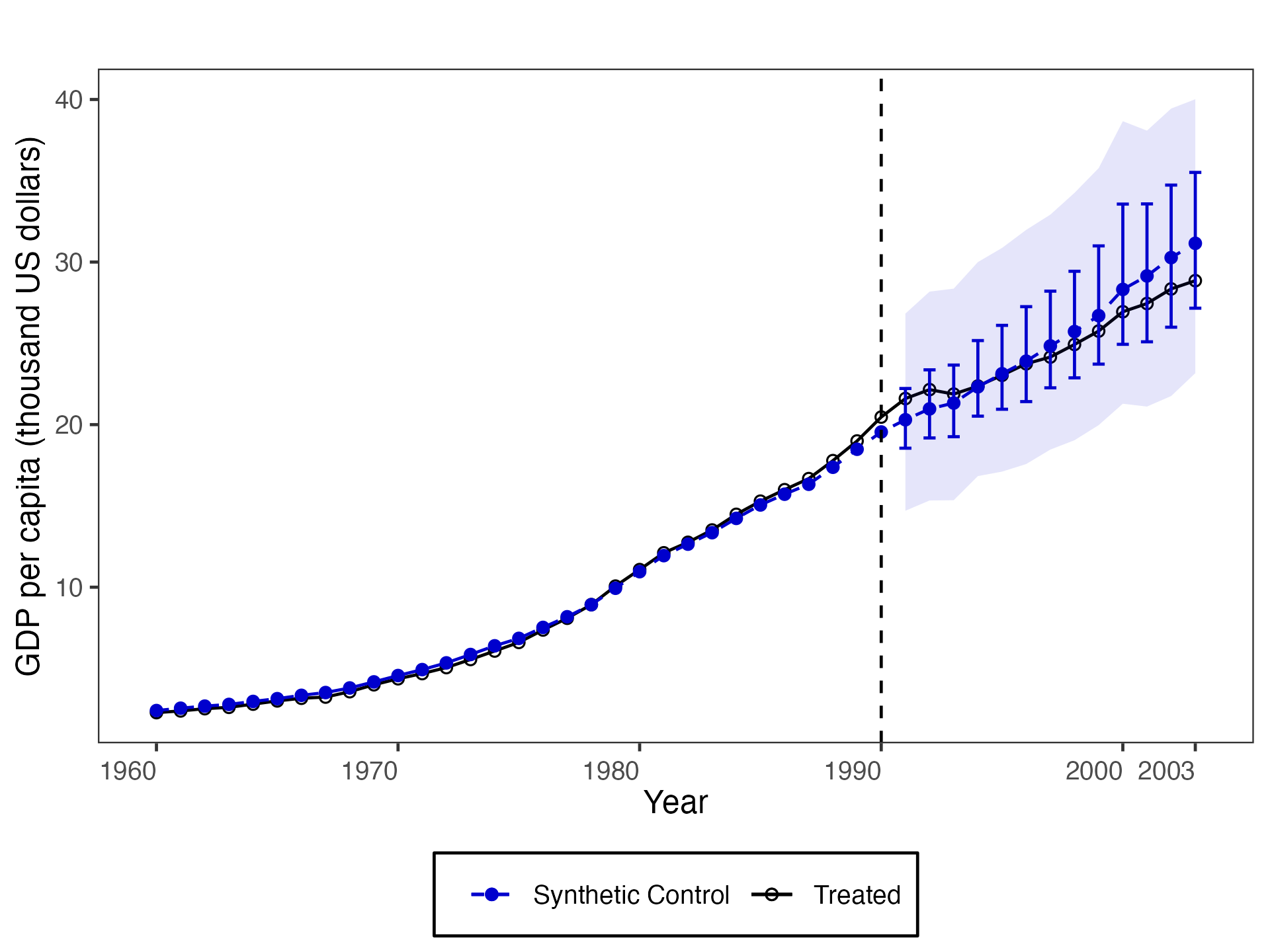}
         \caption{lasso}
     \end{subfigure}
     \\
     \begin{subfigure}[b]{0.48\textwidth}
         \centering
         \includegraphics[width=\textwidth]{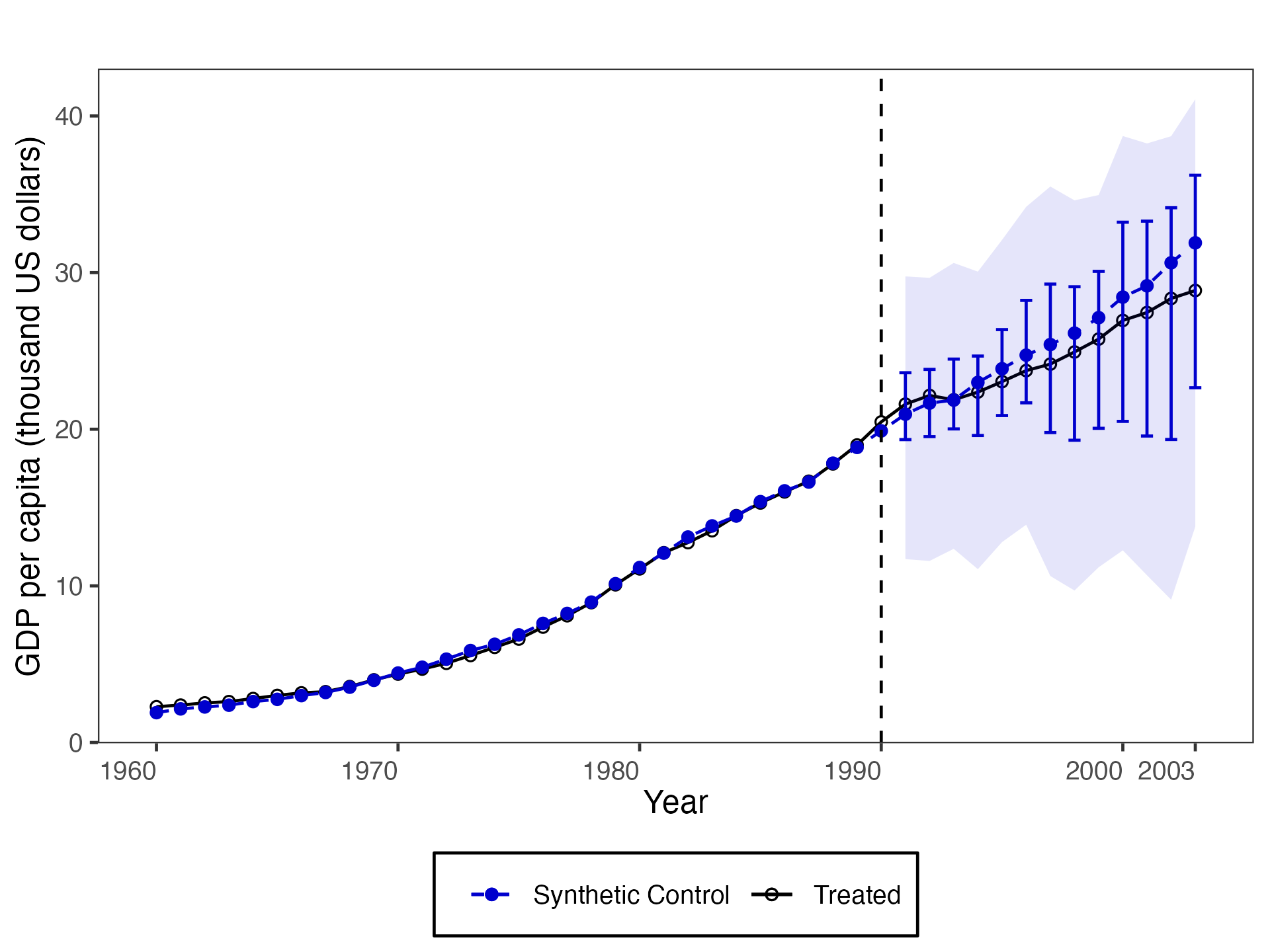}
         \caption{ridge}
     \end{subfigure} \hfill
     \begin{subfigure}[b]{0.48\textwidth}
         \centering
         \includegraphics[width=\textwidth]{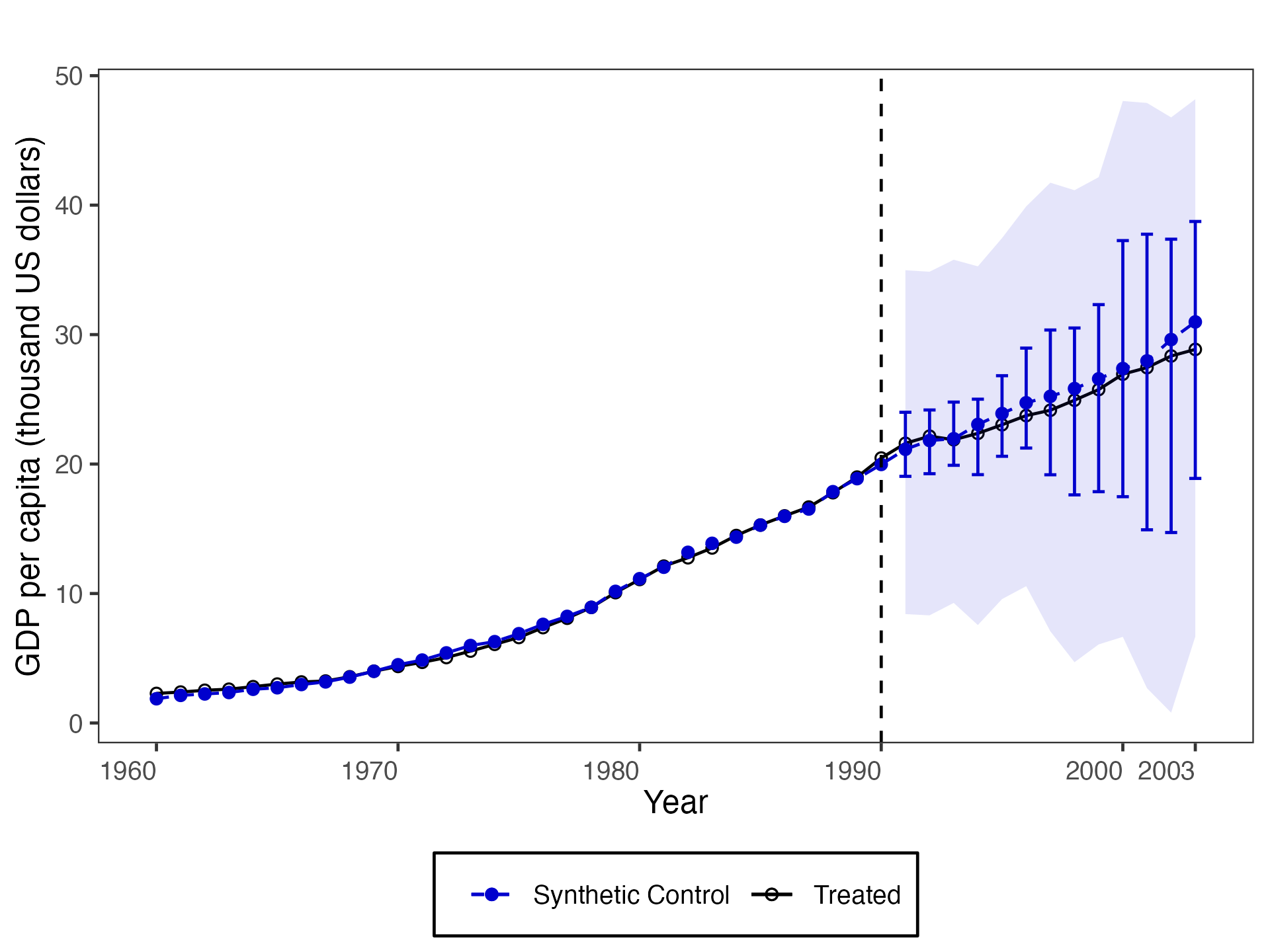}
         \caption{least squares}
     \end{subfigure} \\
     \begin{subfigure}[b]{0.48\textwidth}
         \centering
         \includegraphics[width=\textwidth]{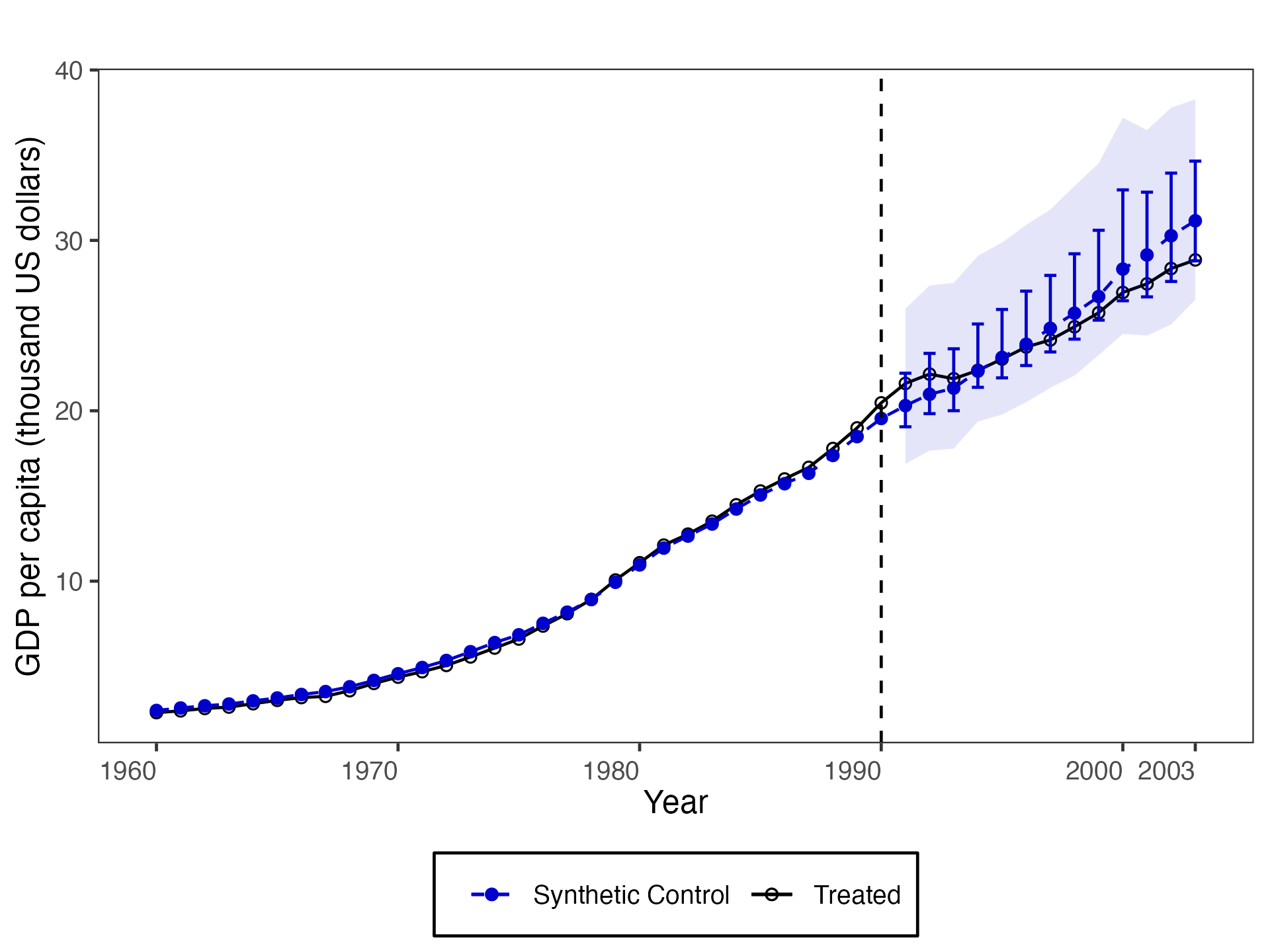}
         \caption{L1-L2}
     \end{subfigure}
  \par
\begin{center}
	\parbox[1]{\textwidth}{\footnotesize \textit{Notes:} The black line shows the level of the outcome for the treated unit, $Y_{1t}(1)$, $t=1963,\ldots, 2003$, whilst the blue line shows the level of the outcome for the synthetic control, $\widehat{Y}_{1t}(0)$, $t=1963,\ldots,2003$. The blue bars report 90\% prediction intervals for $Y_{1t}(0)$. In-sample uncertainty is quantified by means of 1000 simulations of \eqref{eq:in sam obj}, whereas out-of-sample uncertainty is quantified through sub-Gaussian bounds. Blue shaded areas display 90\% simultaneous prediction intervals. In panel (c), $Q=0.903$, whereas in panel (e) $Q=1,Q_2=0.903$.}
\end{center}   
\end{figure}

Results are reported in Figure \ref{fig:multifeatures}, where blue shaded areas depict 90\% simultaneous prediction intervals for periods from 1991 to 2004.

\section{Conclusion}\label{sec:conclusion}

This article introduced the \texttt{R} software package \texttt{scpi}, which implements point estimation/prediction and inference/uncertainty quantification procedures for synthetic control methods. The package is also available in the \texttt{Stata} and \texttt{Python} statistical platforms, as described in the appendices. Further information can be found at \url{https://nppackages.github.io/scpi/}.

\section{Acknowledgments}

We thank Alberto Abadie and Bartolomeo Stellato for many insightful discussions. Cattaneo and Titiunik gratefully acknowledge financial support from the National Science Foundation (SES-2019432), and Cattaneo gratefully acknowledges financial support from the National Institute of Health (R01 GM072611-16).

\bibliography{CFPT_2022_SCPI--bib}
\bibliographystyle{jasa}

\clearpage
\appendix
\section{Appendix: Python Illustration}\label{app:python}
\begin{figure}[H]
     \centering
     \caption{Uncertainty quantification with different types of $\mathcal{W}$ using 90\% prediction intervals.}
     \label{fig:app py}
      \begin{subfigure}[b]{0.33\textwidth}
     \centering
     \includegraphics[width=\textwidth]{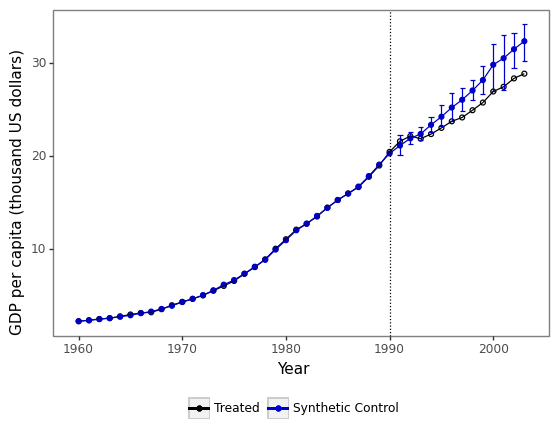}
     \caption{simplex}
 \end{subfigure} \hfill
     \begin{subfigure}[b]{0.33\textwidth}
         \centering
         \includegraphics[width=\textwidth]{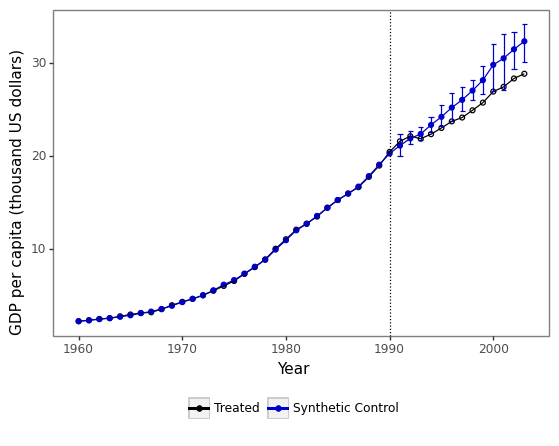}
         \caption{lasso}
     \end{subfigure}\hfill
     \begin{subfigure}[b]{0.33\textwidth}
         \centering
         \includegraphics[width=\textwidth]{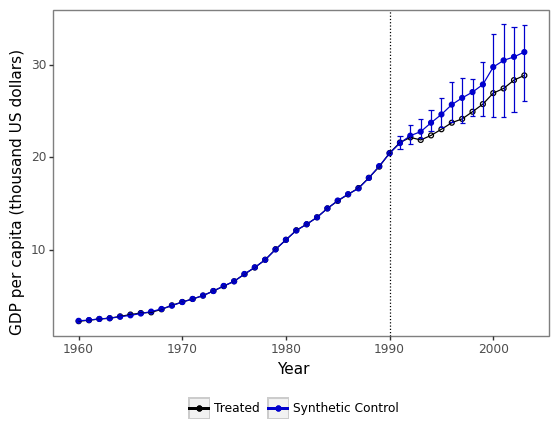}
         \caption{ridge}
     \end{subfigure} \\
     \begin{subfigure}[b]{0.33\textwidth}
         \centering
         \includegraphics[width=\textwidth]{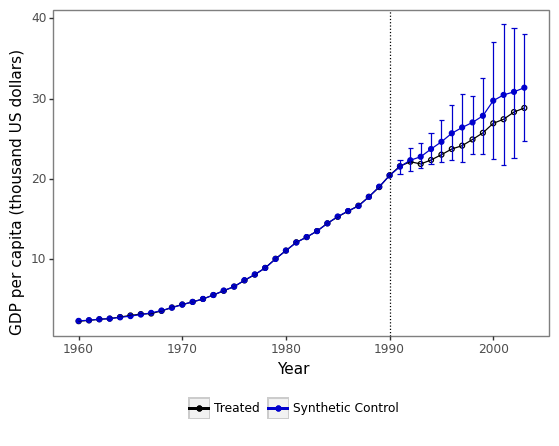}
         \caption{least squares}
     \end{subfigure} 
     \begin{subfigure}[b]{0.33\textwidth}
         \centering
         \includegraphics[width=\textwidth]{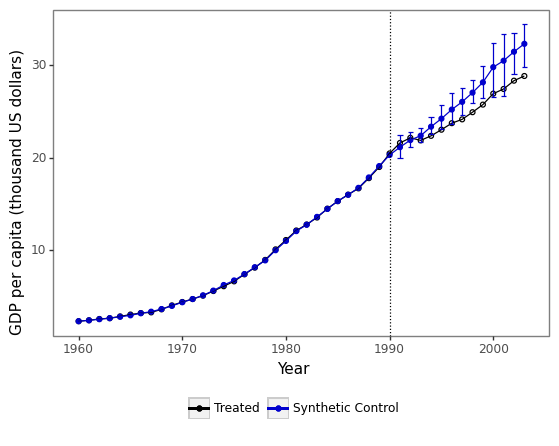}
         \caption{L1-L2}
     \end{subfigure}
  \par
\begin{center}
	\parbox[1]{\textwidth}{\footnotesize \textit{Notes:} The black line shows the level of the outcome for the treated unit, $Y_{1t}(1),t=1963,\ldots, 2003$, whilst the blue line shows the level of the outcome for the synthetic control, $\widehat{Y}_{1t}(0)$, $t=1963,\ldots,2003$. The blue bars report 90\% prediction intervals for $Y_{1t}(0)$. In-sample uncertainty is quantified by means of 1000 simulations of \eqref{eq:in sam obj}, whereas out-of-sample uncertainty is quantified through sub-Gaussian bounds. In panel (c), $Q=0.906$, whereas in panel (e) $Q=1,Q_2=0.906$.}
\end{center}   
\end{figure}

\lstset{language=Python}\lstset{style=Pystyle}
This appendix section shows how to conduct the same analysis carried out in Section \ref{sec: illustration} using the companion \texttt{Python} package. Figure \ref{fig:app py} shows the main results. The L1-L2 constraint is currently not implemented in the \texttt{Python} version of the \texttt{scpi} package due to technical difficulties with the optimizer \texttt{nlopt}. Replication files and data are available at \url{https://nppackages.github.io/scpi/}.

{\singlespacing
\begin{lstlisting}[numbers=none]
###################################################################
# Replication file for Cattaneo, Feng, Palomba, and Titiunik (2022)
###################################################################

########################################
# Load SCPI_PKG package
import pandas
import numpy
import random
import os
from warnings import filterwarnings
from plotnine import ggtitle, ggsave

from scpi_pkg.scdata import scdata
from scpi_pkg.scdataMulti import scdataMulti
from scpi_pkg.scest import scest
from scpi_pkg.scpi import scpi
from scpi_pkg.scplot import scplot
from scpi_pkg.scplotMulti import scplotMulti

os.chdir('YOUR_PATH_HERE')
filterwarnings('ignore')

########################################
# One feature (gdp)
########################################

########################################
# Load database
data = pandas.read_csv('scpi_germany.csv')

########################################
# Set options for data preparation
id_var = 'country'
outcome_var = 'gdp'
time_var = 'year'
period_pre = numpy.arange(1960, 1991)
period_post = numpy.arange(1991, 2004)
unit_tr = 'West Germany'
unit_co = list(set(data[id_var].to_list()))
unit_co = [cou for cou in unit_co if cou != 'West Germany']
constant = True
cointegrated_data = True

data_prep = scdata(df=data, id_var=id_var, time_var=time_var,
                   outcome_var=outcome_var, period_pre=period_pre,
                   period_post=period_post, unit_tr=unit_tr,
                   unit_co=unit_co, cointegrated_data=cointegrated_data,
                   constant=constant)

####################################
# SC - point estimation with simplex
est_si = scest(data_prep, w_constr={'name': 'simplex'})
print(est_si)

####################################
# Set options for inference
w_constr = {'name': 'simplex', 'Q': 1}
u_missp = True
u_sigma = 'HC1'
u_order = 1
u_lags = 0
e_method = 'gaussian'
e_order = 1
e_lags = 0
sims = 1000
cores = 1

for mtd in ['simplex', 'lasso', 'ridge', 'L1-L2', 'ols']:
    if mtd in ['ridge', 'L1-L2']:
        lgapp = 'generalized'
    else:
        lgapp = 'linear'
    random.seed(8894)
    pi_si = scpi(data_prep, sims=sims, w_constr={'name': mtd},
                 u_order=u_order, u_lags=u_lags, e_order=e_order,
                 e_lags=e_lags, e_method=e_method, u_missp=u_missp,
                 lgapp=lgapp, u_sigma=u_sigma, cores=cores)

    plot = scplot(pi_si, x_lab='Year', e_method=e_method,
                  y_lab='GDP per capita (thousand US dollars)')

    plot = plot + ggtitle('')
    pltname = 'py_germany_unc_' + str(mtd) + '.png'
    ggsave(filename=pltname, plot=plot)


########################################
# Multiple features (gdp, trade)
########################################

########################################
# Load database
data = pandas.read_csv('scpi_germany.csv')

########################################
# Set options for data preparation
id_var = 'country'
outcome_var = 'gdp'
time_var = 'year'
period_pre = numpy.arange(1960, 1991)
period_post = numpy.arange(1991, 2004)
unit_tr = 'West Germany'
unit_co = list(set(data[id_var].to_list()))
unit_co = [cou for cou in unit_co if cou != 'West Germany']
constant = False
cointegrated_data = True
cov_adj = [['constant'], ['constant']]

data_prep = scdata(df=data, id_var=id_var, time_var=time_var,
                   outcome_var=outcome_var, period_pre=period_pre,
                   period_post=period_post, unit_tr=unit_tr, constant=constant,
                   unit_co=unit_co, cointegrated_data=cointegrated_data,
                   features=['gdp', 'trade'], cov_adj=cov_adj)

####################################
# SC - point estimation with simplex
est_si = scest(data_prep, w_constr={'name': 'simplex'})
print(est_si)

####################################
# Set options for inference
w_constr = {'name': 'simplex', 'Q': 1}
u_missp = True
u_sigma = 'HC1'
u_order = 1
u_lags = 0
e_method = 'gaussian'
e_order = 1
e_lags = 0
sims = 1000
cores = 1

for mtd in ['simplex', 'lasso', 'ridge', 'L1-L2', 'ols']:
    if mtd in ['ridge', 'L1-L2']:
        lgapp = 'generalized'
    else:
        lgapp = 'linear'

    random.seed(8894)
    pi_si = scpi(data_prep, sims=sims, w_constr={'name': mtd},
                 u_order=u_order, u_lags=u_lags, e_order=e_order,
                 e_lags=e_lags, e_method=e_method, u_missp=u_missp,
                 lgapp=lgapp, u_sigma=u_sigma, cores=cores)

    plot = scplot(pi_si, x_lab='Year', e_method=e_method,
                  y_lab='GDP per capita (thousand US dollars)')

    plot = plot + ggtitle('')
    pltname = 'py_germany_unc_' + str(mtd) + '.png'
    ggsave(filename=pltname, plot=plot)
\end{lstlisting}}

\clearpage

\uline{\textit{Case I}: $M=1$}
\clearpage

\uline{\textit{Case II}: $M=2$}

\begin{figure}[H]
     \centering
     \caption{Uncertainty quantification with different types of $\mathcal{W}$ using 90\% prediction intervals.}
     \label{fig:app py multi}
      \begin{subfigure}[b]{0.33\textwidth}
     \centering
     \includegraphics[width=\textwidth]{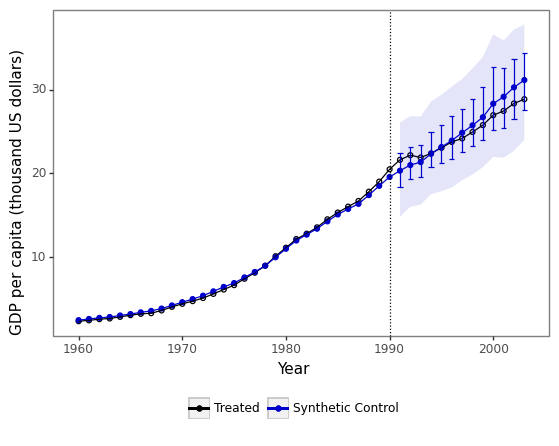}
     \caption{simplex}
 \end{subfigure} \hfill
     \begin{subfigure}[b]{0.33\textwidth}
         \centering
         \includegraphics[width=\textwidth]{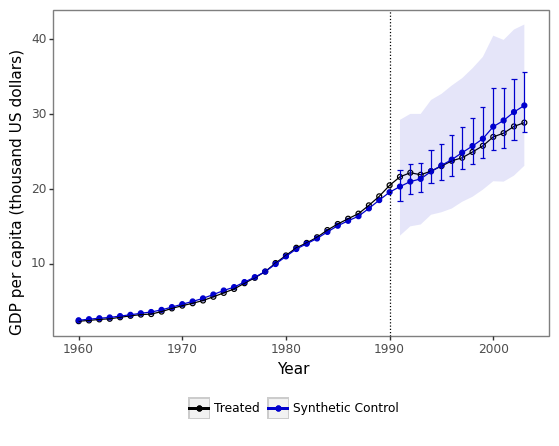}
         \caption{lasso}
     \end{subfigure}\hfill
     \begin{subfigure}[b]{0.33\textwidth}
         \centering
         \includegraphics[width=\textwidth]{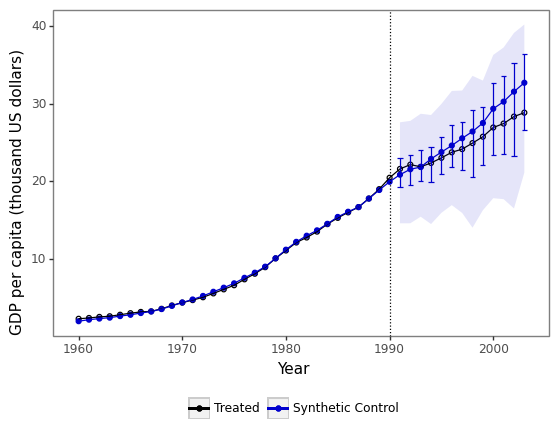}
         \caption{ridge}
     \end{subfigure} \\
     \begin{subfigure}[b]{0.33\textwidth}
         \centering
         \includegraphics[width=\textwidth]{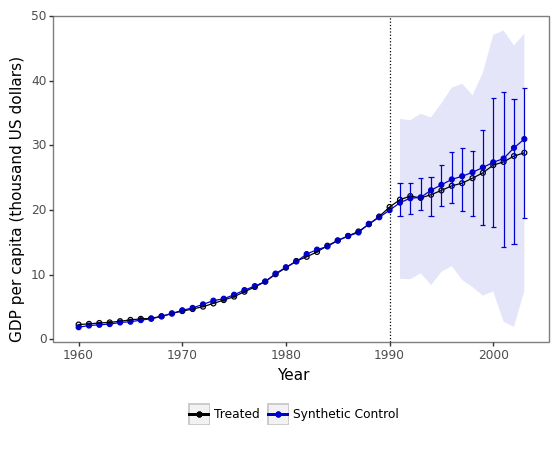}
         \caption{least squares}
     \end{subfigure} 
     \begin{subfigure}[b]{0.33\textwidth}
         \centering
         \includegraphics[width=\textwidth]{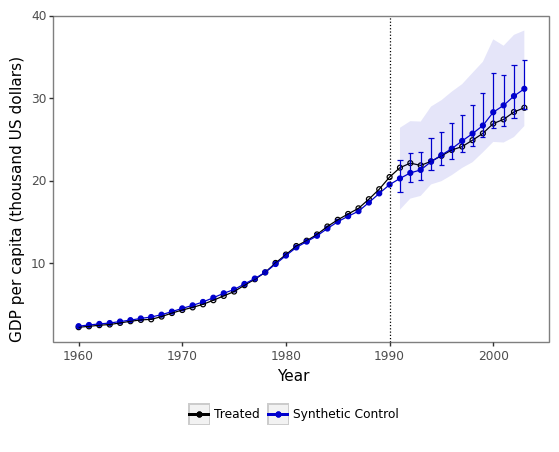}
         \caption{L1-L2}
     \end{subfigure}     
  \par
\begin{center}
	\parbox[1]{\textwidth}{\footnotesize \textit{Notes:} The black line shows the level of the outcome for the treated unit, $Y_{1t}(1)$, $t=1963,\ldots, 2003$, whilst the blue line shows the level of the outcome for the synthetic control, $\widehat{Y}_{1t}(0)$, $t=1963,\ldots,2003$. The blue bars report 90\% prediction intervals for $Y_{1t}(0)$. In-sample uncertainty is quantified by means of 1000 simulations of \eqref{eq:in sam obj}, whereas out-of-sample uncertainty is quantified through sub-Gaussian bounds. Blue shaded areas display 90\% simultaneous prediction intervals. In panel (c), $Q=0.903$, whereas in panel (e) $Q=1,Q_2=0.903$.}
\end{center}   
\end{figure}

\clearpage

\section{Appendix: Stata Illustration}\label{app:stata}
\begin{figure}[H]
     \centering
     \caption{Uncertainty quantification with different types of $\mathcal{W}$ using 90\% prediction intervals.}
     \label{fig:app stata}
     \begin{subfigure}[b]{0.33\textwidth}
         \centering
         \includegraphics[width=\textwidth]{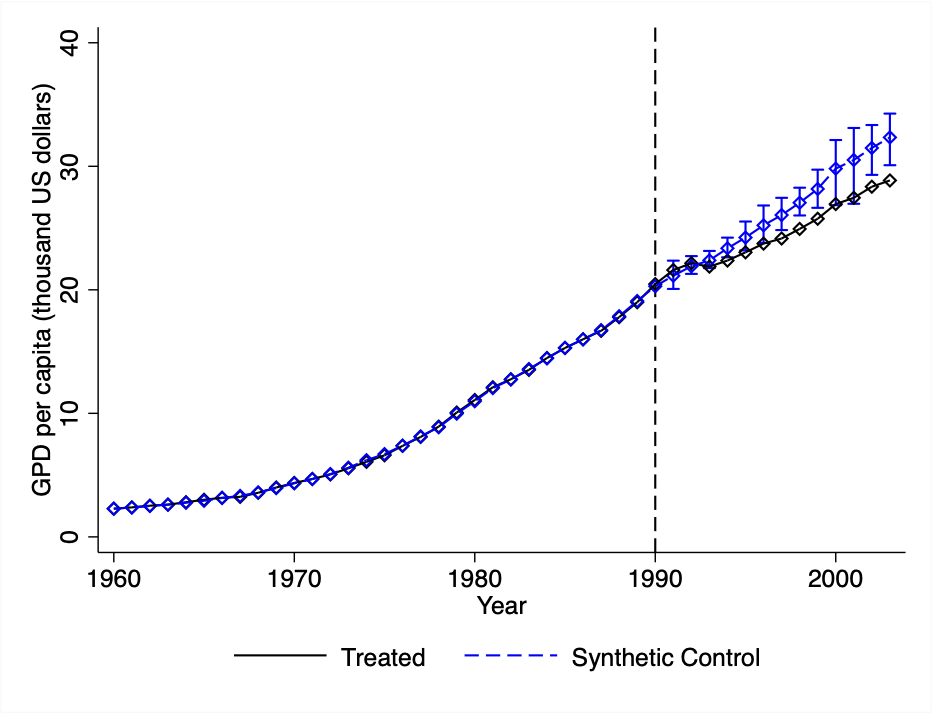}
         \caption{simplex}
     \end{subfigure}\hfill
     \begin{subfigure}[b]{0.33\textwidth}
         \centering
         \includegraphics[width=\textwidth]{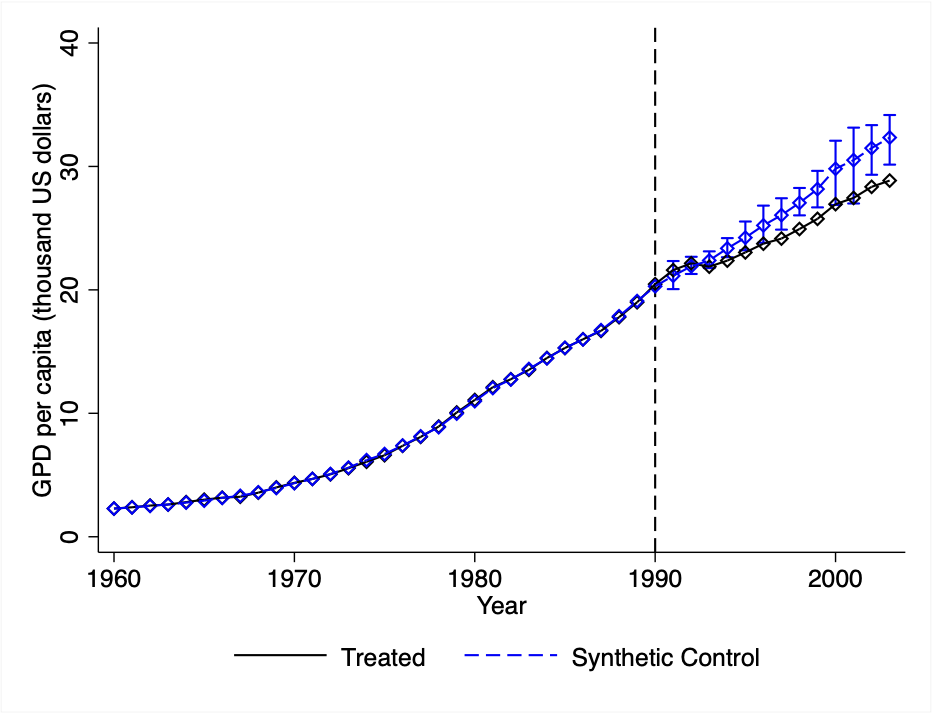}
         \caption{lasso}
     \end{subfigure}\hfill
     \begin{subfigure}[b]{0.33\textwidth}
         \centering
         \includegraphics[width=\textwidth]{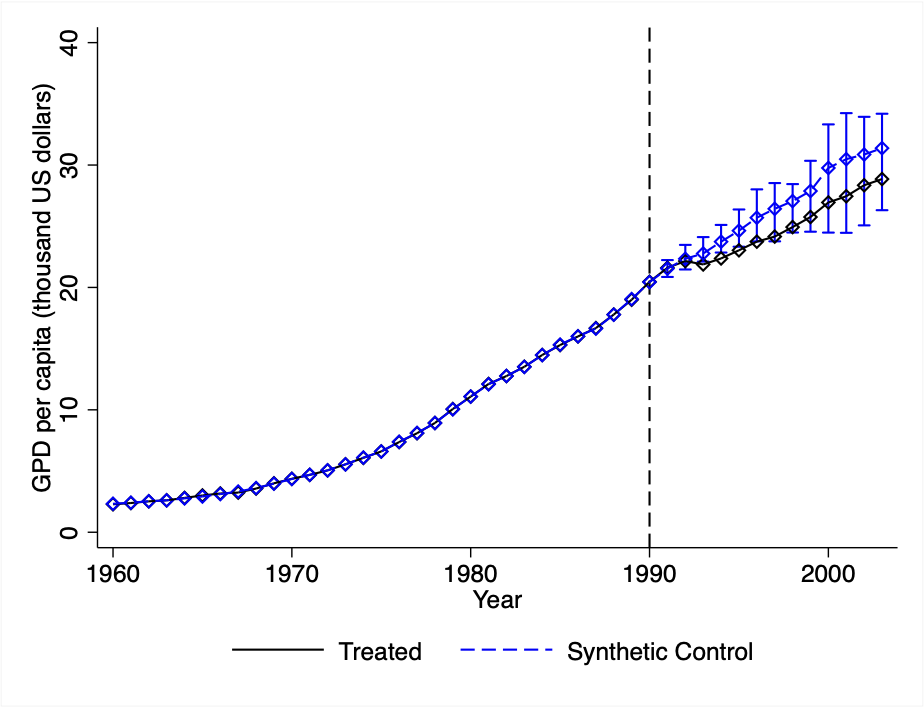}
         \caption{ridge}
     \end{subfigure} \\
     \begin{subfigure}[b]{0.33\textwidth}
         \centering
         \includegraphics[width=\textwidth]{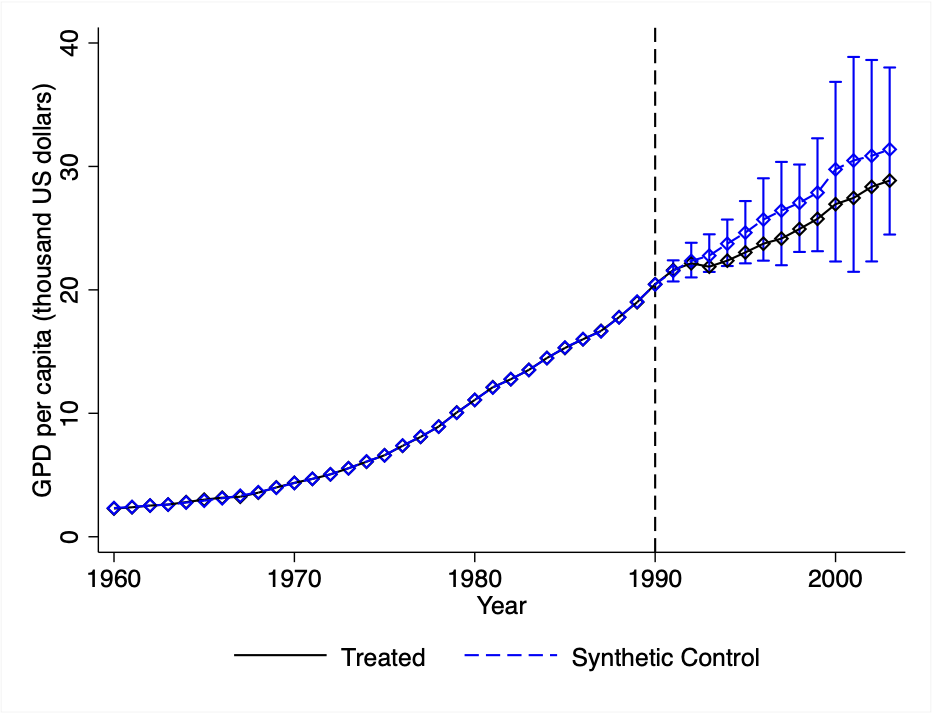}
         \caption{least squares}
     \end{subfigure} 
     \begin{subfigure}[b]{0.33\textwidth}
         \centering
         \includegraphics[width=\textwidth]{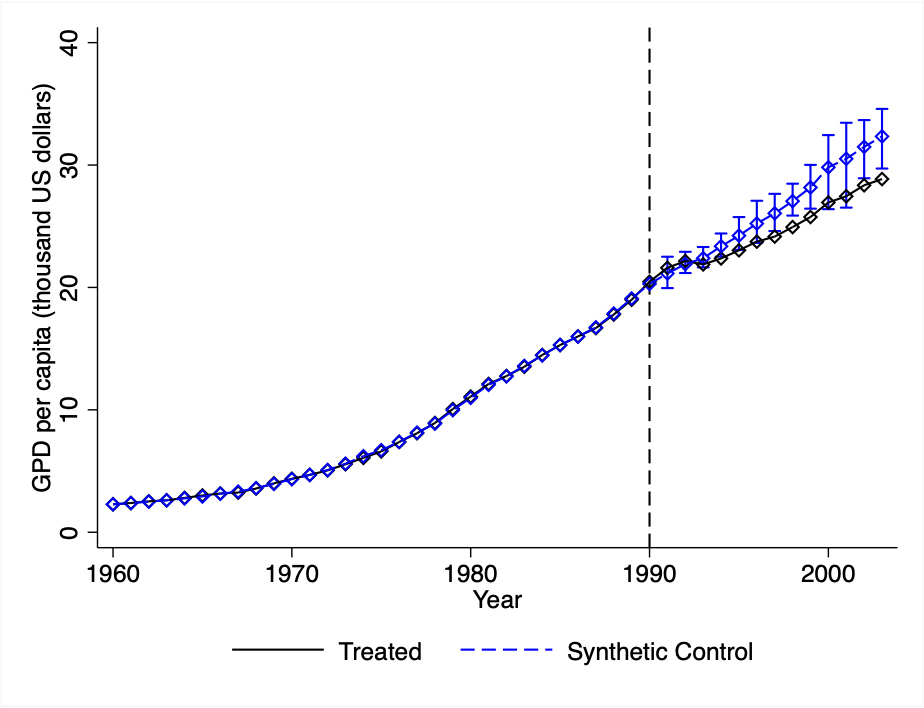}
         \caption{L1-L2}
     \end{subfigure}     
  \par
\begin{center}
	\parbox[1]{\textwidth}{\footnotesize \textit{Notes:} The black line shows the level of the outcome for the treated unit, $Y_{1t}(1)$, $t=1963,\ldots, 2003$, whilst the blue line shows the level of the outcome for the synthetic control, $\widehat{Y}_{1t}(0)$, $t=1963,\ldots,2003$. The blue bars report 90\% prediction intervals for $Y_{1t}(0)$. In-sample uncertainty is quantified by means of 1000 simulations of \eqref{eq:in sam obj}, whereas out-of-sample uncertainty is quantified through sub-Gaussian bounds. In panel (c), $Q=0.906$, whereas in panel (e) $Q=1,Q_2=0.906$.}
\end{center}   
\end{figure}

\lstset{language=stata}\lstset{style=stata-editor}
This appendix section replicates the analysis conducted in Section \ref{sec: illustration} for $M=1$ using the companion \texttt{Stata} package. Main results are shown in Figure \ref{fig:app stata}. The L1-L2 constraint is currently not implemented in the \texttt{Stata} version of the \texttt{scpi} package due to technical difficulties with the optimizer \texttt{nlopt}. Replication files and data are available at \url{https://nppackages.github.io/scpi/}.

{\singlespacing
\begin{lstlisting}[language = Stata, numbers=none]
******************************************************************
* Replication file - Cattaneo, Feng, Palomba, and Titiunik (2022)
******************************************************************

* Load dataset
use "scpi_germany.dta", clear

***********************************************
** One feature (gdp)
***********************************************

* Prepare data
scdata gdp, dfname("python_scdata") id(country) outcome(gdp) time(year) ///
			treatment(status) cointegrated constant 

* Quantify uncertainty
local lgapp "linear"
foreach method in "simplex" "lasso" "ols" "ridge" "L1-L2" {
	if inlist("`method'", "ridge", "L1-L2") {
		local lgapp "generalized"
	} 
	set seed 8894
	scpi, dfname("python_scdata") name(`method') e_method(gaussian) u_missp ///
		  lgapp("`lgapp'") sims(1000) 

	scplot, uncertainty("gaussian") gphoptions(note("") xtitle("Year") ///
		ytitle("GPD per capita (thousand US dollars)")) 
	graph export "stata_germany_unc_`method'.png", replace
}

***********************************************
** Multiple features (gdp, trade)
***********************************************

* Prepare data
scdata gdp trade, dfname("python_scdata") id(country) outcome(gdp) time(year) ///
			treatment(status) cointegrated  covadj("constant")

* Quantify uncertainty
local lgapp "linear"
foreach method in "simplex" "lasso" "ols" "ridge" "L1-L2" {
	if inlist("`method'", "ridge", "L1-L2") {
		local lgapp "generalized"
	} 
	
	set seed 8894
	scpi, dfname("python_scdata") name(`method') e_method(gaussian) u_missp ///
		  lgapp("`lgapp'") sims(1000) 

	scplot, uncertainty("gaussian") gphoptions(note("") xtitle("Year") ///
		ytitle("GPD per capita (thousand US dollars)")) joint 
	graph export "stata_germany_unc_`method'_multi.png", replace
}
\end{lstlisting}}

\uline{\textit{Case I}: $M=1$}
\clearpage

\uline{\textit{Case II}: $M=2$}

\begin{figure}[H]
     \centering
     \caption{Uncertainty quantification with different types of $\mathcal{W}$ using 90\% prediction intervals.}
     \label{fig:app stata multi}
     \begin{subfigure}[b]{0.33\textwidth}
         \centering
         \includegraphics[width=\textwidth]{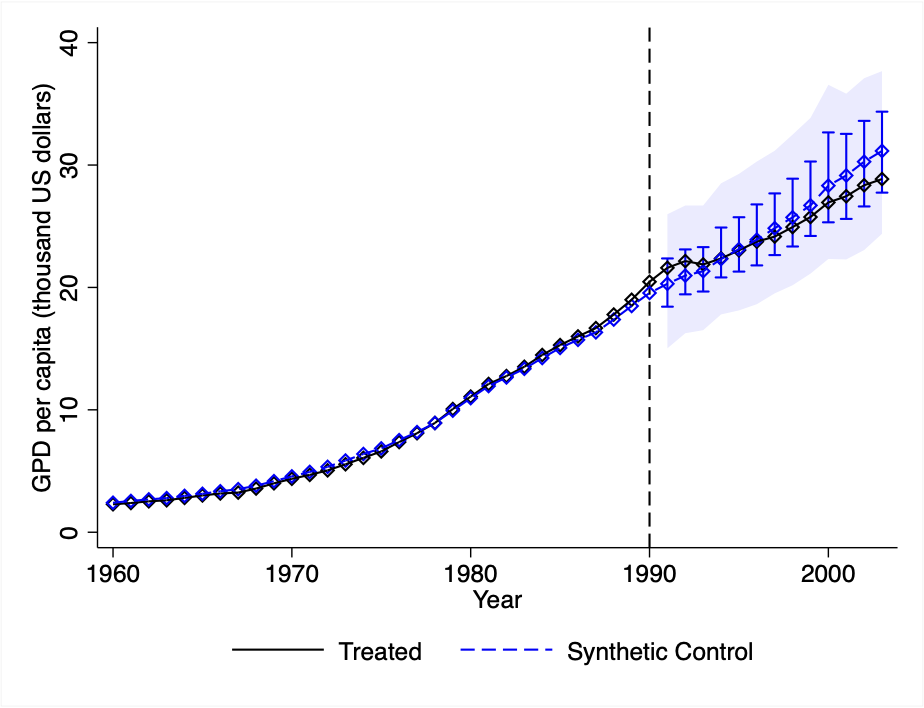}
         \caption{simplex}
     \end{subfigure}\hfill
     \begin{subfigure}[b]{0.33\textwidth}
         \centering
         \includegraphics[width=\textwidth]{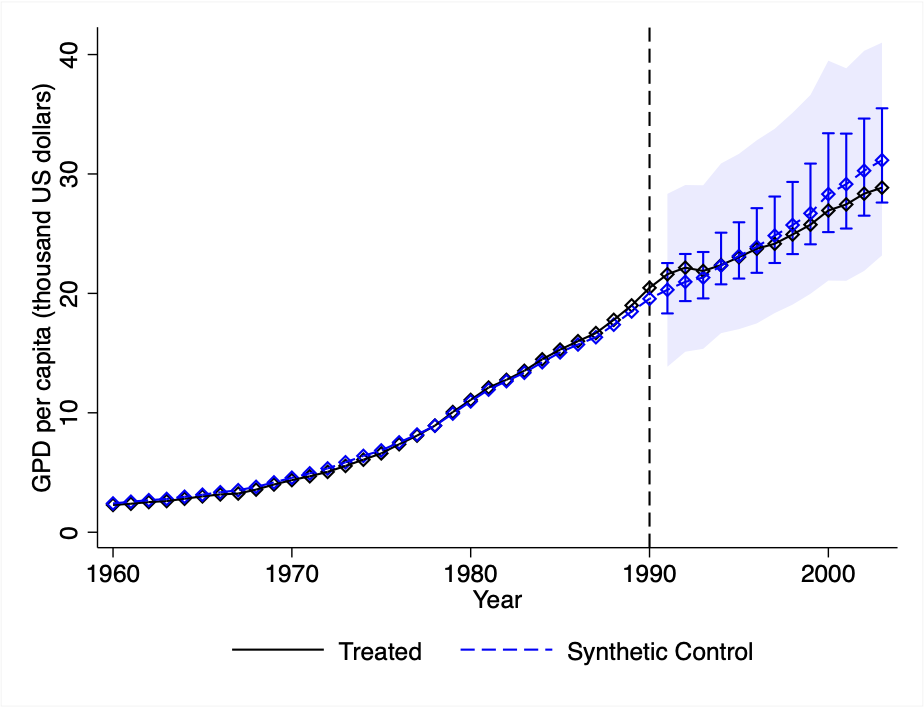}
         \caption{lasso}
     \end{subfigure}\hfill
     \begin{subfigure}[b]{0.33\textwidth}
         \centering
         \includegraphics[width=\textwidth]{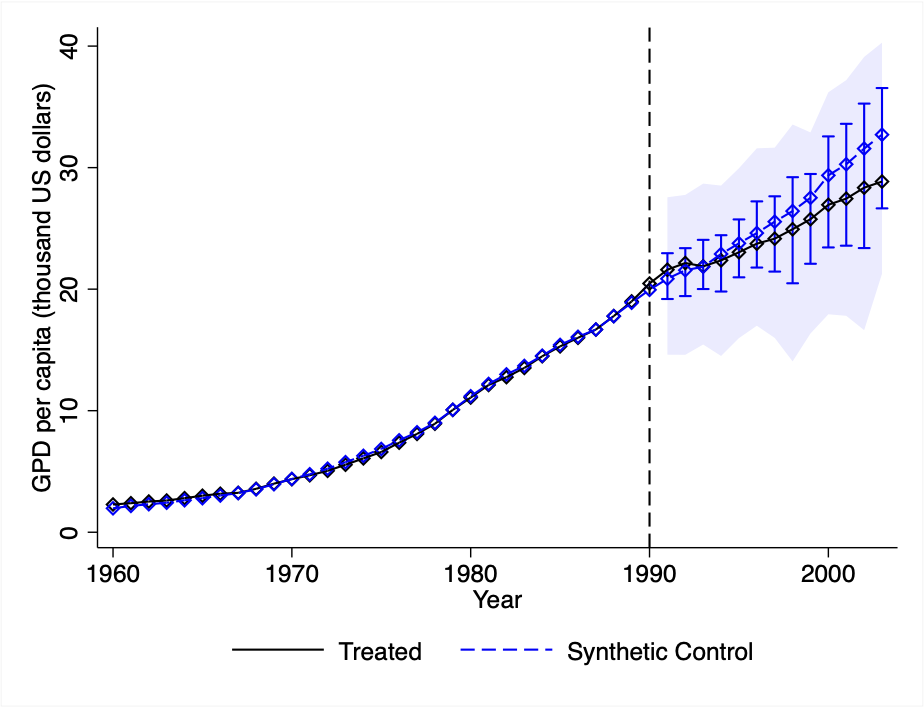}
         \caption{ridge}
     \end{subfigure} \\
     \begin{subfigure}[b]{0.33\textwidth}
         \centering
         \includegraphics[width=\textwidth]{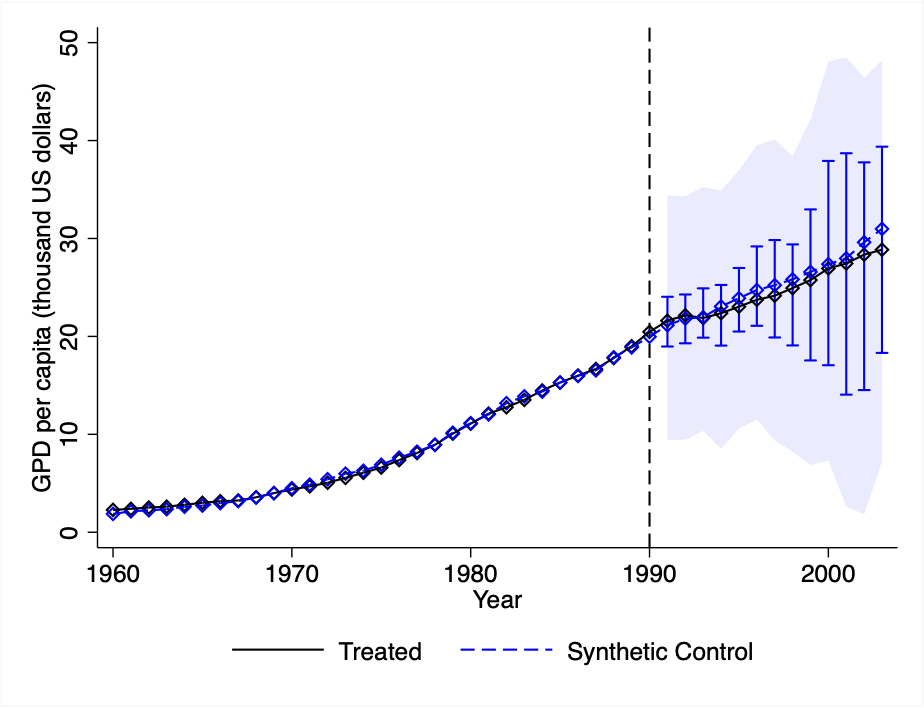}
         \caption{least squares}
     \end{subfigure} 
     \begin{subfigure}[b]{0.33\textwidth}
         \centering
         \includegraphics[width=\textwidth]{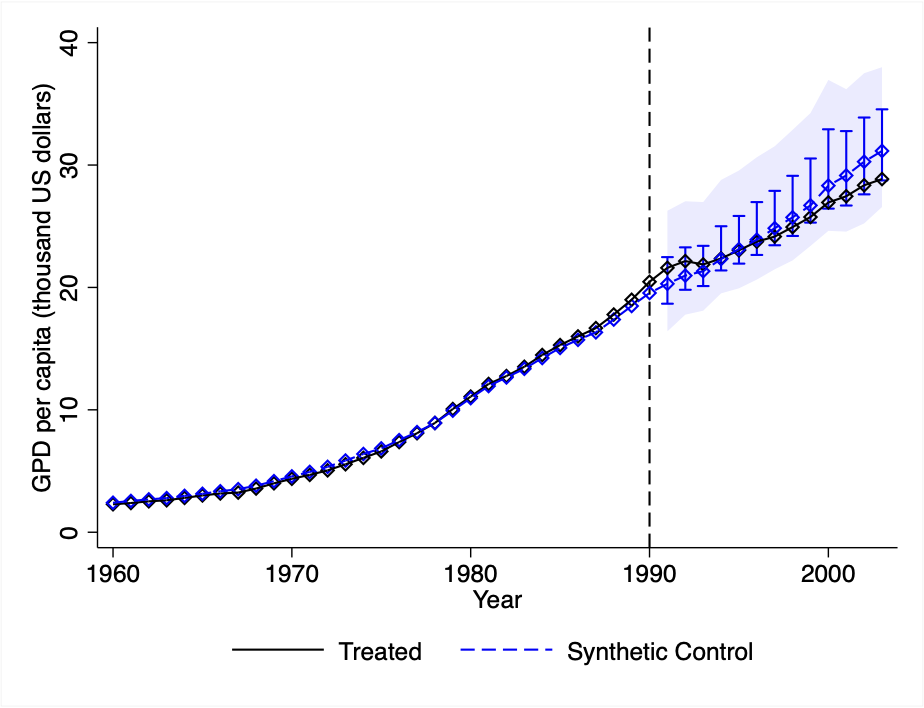}
         \caption{L1-L2}
     \end{subfigure}          
  \par
\begin{center}
	\parbox[1]{\textwidth}{\footnotesize \textit{Notes:} The black line shows the level of the outcome for the treated unit, $Y_{1t}(1)$, $t=1963,\ldots, 2003$, whilst the blue line shows the level of the outcome for the synthetic control, $\widehat{Y}_{1t}(0)$, $t=1963,\ldots,2003$. The blue bars report 90\% prediction intervals for $Y_{1t}(0)$. In-sample uncertainty is quantified by means of 1000 simulations of \eqref{eq:in sam obj}, whereas out-of-sample uncertainty is quantified through sub-Gaussian bounds. Blue shaded areas display 90\% simultaneous prediction intervals. In panel (c), $Q=0.903$, whereas in panel (e) $Q=1,Q_2=0.903$.}
\end{center}   
\end{figure}
\end{document}